\documentclass[11pt,a4paper]{article}
\usepackage{jcappub}
\usepackage{bm}
\usepackage{amsmath}
\usepackage{latexsym,amsmath,amssymb,amsfonts,mathrsfs}
\usepackage{appendix}
\usepackage[usenames,dvipsnames]{xcolor}
\usepackage{verbatim}
\usepackage{ulem}

\def\dd{\mathrm{d}}
\def\mcA{\mathcal{A}}
\def\mcP{\mathcal{P}}
\def\em{{\rm em}}
\def\inf{{\rm inf}}
\def\tot{{\rm tot}}

\def\Mpl{M_{\rm Pl}}
\def\GeV{{\rm GeV}}
\def\Mpc{{\rm Mpc}}

\def\G{{\rm G}}

\def\osc{{\rm osc}}
\def\BD{{\rm BD}}

\definecolor{DGreen}{rgb}{0,0.7,0}

\title{
Scale-invariant Helical Magnetic Fields
from Inflation 
}
\author[a,b]{Tomohiro Fujita}
\author[b]{, Ruth Durrer}

\affiliation[a]{Department of Physics, Kyoto University, Kyoto, 606-8502, Japan}
\affiliation[b]{D\'epartment de Physique Th\'eorique and Center for Astroparticle Physics, \\
Universit\'e de Gen\`eve, Quai E.
Ansermet 24, CH-1211 Gen\`eve 4, Switzerland}

\emailAdd{t.fujita@tap.scphys.kyoto-u.ac.jp}
\emailAdd{ruth.durrer@unige.ch}
\abstract{
We discuss a model which can generate scale-invariant helical magnetic fields
on large scales ($\lesssim 1$Mpc) in the primordial universe.
It is also shown that the electric conductivity becomes significant and terminates magnetogenesis even before reheating is completed.
By solving the electromagnetic dynamics taking  conductivity into account, we find that magnetic fields with amplitude $B\simeq 10^{-15}{\rm G}$ at present can be generated without encountering a backreaction or strong coupling problem. }

\arxivnumber{1904.XXXXX}

\usepackage{color}

\begin{document}


\maketitle

\section{Introduction}

It is well known that galaxies and galactic clusters have magnetic fields with strength of the order of $10^{-6}\G$~\cite{Bernet:2008qp,Bonafede:2010xg,Feretti:2012vk}.
However, the origin of these fields is not understood.
One possibility is that magnetic fields generated in the early universe are stretched over cosmological scales and  serve as the seed for the galactic and cluster magnetic fields. This primordial magnetogenesis scenario predicts that weaker magnetic fields exist also in inter-galactic regions, voids.
Indeed, a lower bound on cosmological magnetic fields permeating the inter-galactic medium has been derived from blazar observation as~\cite{Neronov:1900zz, Dolag:2010ni, Essey:2010nd, Taylor:2011bn, Chen:2014rsa,Finke:2015ona,Biteau:2018tmv} 
\begin{equation}
B\gtrsim 10^{-16}\G \times\left\{\begin{array}{cc}
1 & (\lambda_B\gtrsim 1\Mpc) \\
\sqrt{1\Mpc/\lambda_B}\quad & (\lambda_B\lesssim 1\Mpc) \\
\end{array}\right.,
\label{Blazer bound}
\end{equation}
where $\lambda_B$ denotes the correlation length of the magnetic fields which is degenerated with the strength $B$ in this constraint.
This observational bound strongly motivates us to seek a mechanism to generate magnetic fields in the primordial universe.
In addition,  CMB observations put an upper bound on large-scale magnetic fields of about, $B\lesssim 10^{-9}\G$~\cite{Ade:2015cva}.
It is also inferred from diffuse gamma-ray observation that the intergalactic magnetic fields are helical~\cite{Tashiro:2013bxa, Tashiro:2013ita,Chen:2014qva}.
For more details, interested readers are referred to review articles~\cite{Durrer:2013pga,Subramanian:2015lua}.

One of the most studied models of primordial magnetogenesis is the so-called $I^2 FF$ model (or Ratra-model)~\cite{Ratra:1991bn}. This model assumes a scalar field $\Phi$ coupled to the electromagnetic kinetic term as $I^2(\Phi)F_{\mu\nu}F^{\mu\nu}$
and the electromagnetic fields are generated while $I(\Phi)$ evolves during inflation. However, several issues of this model have been pointed out~\cite{Gasperini:1995dh,Demozzi:2009fu,Fujita:2012rb}. Among them, the back reaction problem, which involves the IR divergence of the energy density of the electric field, often becomes relevant~\cite{Demozzi:2009fu,Bamba:2003av,Kanno:2009ei}.
One should also be concerned with the curvature perturbation induced by the generated electromagnetic fields which must be consistent with the CMB observation~\cite{Barnaby:2012tk,Giovannini:2013rme,Fujita:2013pgp,Fujita:2014sna}.
Addressing these issues, Ferreira et al.~\cite{Ferreira:2013sqa,Ferreira:2014hma} proposed to introduce the onset of the evolution of $I(\Phi)$ during observable inflation which works as IR cut-off. Kobayashi also proposed to extend the $I(\Phi)$ evolution even after inflation which enables further amplification of the electromagnetic fields~\cite{Kobayashi:2014sga}.
Combining these proposals, the $I^2 FF$ model can produce $B\simeq10^{-15}\G$ at present, but the generated magnetic fields are non-helical~\cite{Fujita:2016qab}.

Another well-studied class of inflationary magnetogenesis models couples a (pseudo-)scalar field $\varphi$  to the $U(1)$ Chern-Simons term, $\varphi F_{\mu\nu}\tilde{F}^{\mu\nu}$~\cite{Turner:1987bw,Garretson:1992vt,Field:1998hi,Durrer:2010mq}.
The pioneering work done by Amber and Sorbo found an analytic solution for the electromagnetic fields during the slow-roll regime of inflation~\cite{Anber:2006xt}.
Later, a comprehensive analysis numerically showed that magnetic field is substantially amplified around the end of inflation~\cite{Fujita:2015iga} which was confirmed at non-linear level by lattice simulation~\cite{Adshead:2016iae}.
Since the magnetic fields generated in this model are helical, 
their correlation length grows after inflation by virtue of an inverse cascade~\cite{Son:1998my,Christensson:2000sp,Kahniashvili:2012uj}.
Recently,  Caprini and Sorbo~\cite{Caprini:2014mja} proposed a hybrid model which contains both  couplings, $I^2 FF$ and $I^2 F\tilde{F}$.
The hybrid model can produce a blue-tilted magnetic field whose current amplitude at $1\Mpc$ scale is $B\lesssim 10^{-21}\G$, but the peak amplitude reaches $B\sim 10^{-14}\G$ at the scale of $\lambda_B\sim 1$pc which roughly satisfies eq.~\eqref{Blazer bound}~\cite{Caprini:2017vnn}.
However, these original works on the hybrid model made the following two assumptions: 
(i) Although a scalar field which drives $I^2$ is not the inflaton and not coupled to the inflaton, it stops rolling at (or slightly before) the end of inflation. (ii) The electric conductivity is negligible during inflation, though reheating instantaneously completes at the end of inflation.

The electric conductivity $\sigma_c$, of the universe plays a crucial role in magnetogenesis scenarios. If $\sigma_c$ is very large, the electric field vanishes and the magnetic field is frozen. Therefore, it is difficult to generate electromagnetic fields once $\sigma_c$ is high and our universe always has a very high conductivity after reheating~\cite{Turner:1987bw}.
This is a good reason to consider magnetogenesis models prior to reheating.
Even in that case, it is the electric conductivity that converts the generated electromagnetic modes into a frozen magnetic field which obeys adiabatic dilution (or undergoes an inverse cascade).
Despite its importance, only a few previous works have highlighted the role of electric conductivity in primordial magnetogenesis~\cite{Turner:1987bw,Martin:2007ue}.
Note that $\sigma_c$ may not be negligible even before the reheating is completed, because it takes a finite time for $\sigma_c$ to grow and affect the growth of electromagnetic fields during reheating.  

In this paper, we extend the hybrid model proposed by Caprini and Sorbo and solve the dynamics of the electromagnetic fields taking into account the electric conductivity.
Regarding the kinetic function $I^2$ driven by a spectator field, 
we consider that $I$ starts decreasing during inflation and continues evolving even after inflation.
As a result, scale-invariant helical magnetic fields strong enough to explain the blazar bound eq.~\eqref{Blazer bound} can be generated. Furthermore, we explicitly show how the electric conductivity terminates the amplification of the electromagnetic fields.
We also find that the conductivity stops magnetogenesis before $I$ stops evolving and well before the reheating completion.
With our fiducial parameter, a magnetic field strength of $B\simeq 10^{-14}\G$ with a coherence length of $\lambda_B\simeq 1\Mpc$ at present can be generated.

This paper is organized as follows. 
In section~\ref{Magnetogenesis Model}, we describe the model setup and 
obtain the electromagnetic spectra by solving the equation of motion without electric conductivity.
Then, the electric conductivity is introduced and  its evolution and its effect on the electromagnetic dynamics are studied in section \ref{Electric Conductivity}. The resulting strength of the generated magnetic field is calculated in section \ref{Magnetic Fields Today}.
The final section \ref{Summary and Discussion} is devoted to the discussion of our results and a summary. 

\section{The Magnetogenesis Model}
\label{Magnetogenesis Model}

\subsection{The model setup}

We consider the following model Lagrangian proposed in Ref.~\cite{Caprini:2014mja}:
\begin{equation}
\mathcal{L}=-\frac{1}{2}(\partial_\mu\phi)^2-V(\phi)
-\frac{1}{2}(\partial_\mu\chi)^2-U(\chi)
-\frac{1}{4}I^2(\chi)\left(F_{\mu\nu}F^{\mu\nu}
-\gamma F_{\mu\nu}\tilde{F}^{\mu\nu}\right),
\label{model lagrangian}
\end{equation}
where $\phi$ is the inflaton, $\chi$ is a spectator scalar field,
$F_{\mu\nu}=\partial_\mu A_\nu-\partial_\nu A_\mu$ is the field strength
of a $U(1)$ gauge field, and $\tilde{F}^{\mu\nu}=\epsilon^{\mu\nu\rho\lambda}F_{\rho\lambda}/(2\sqrt{-g})$
is its dual with the totally antisymmetric tensor $\epsilon^{\mu\nu\rho\lambda}$.
$V(\phi)$ and $U(\chi)$ are the potential of $\phi$ and $\chi$, respectively and
$\gamma$ is a constant.
Although $\chi$ is assumed to have a non-zero background value which drives the evolution of $I(\chi)$, its energy density is always negligible compared to the total energy density and hence $\chi$ is called a spectator.
In this model, the kinetic energy of the $\chi$ field is transferred to the gauge field through the kinetic coupling $I(\chi)$ and  electromagnetic
fields are produced when $I(\chi)$ is varying in time.
The dynamics of this model during inflation has been studied in Ref.~\cite{Caprini:2014mja,Caprini:2017vnn}.
They assumed that $I(\chi)$ varies only during inflation.
However, since $\chi$ is not the inflaton and not coupled to it, we do not have an apparent reason to expect that $I(\chi)$ stops evolving at the end of inflation.
Thus, in this paper, we further investigate this model by considering a post-inflationary dynamics of $I(\chi)$.
Note that the last term in eq.~\eqref{model lagrangian} explicitly breaks the parity symmetry, while the parity violation becomes invisible once $I(\chi)$ stops evolving since the last term is a surface term when $I^2(\chi)=$ constant.

Let us study the dynamics of the gauge field. 
In this paper, we work in Coulomb gauge, $A_0 = \partial_i A_i = 0$.
We decompose and quantize the gauge field as
\begin{equation}
 A_i(t, \bm{x})
 = 
 \sum_{\lambda=\pm} \int \frac{{\rm d}^3 k}{(2\pi)^3} 
 e^{i \bm{k \cdot x}} e_{i}^{(\lambda)}(\hat{\bm{k}}) 
 \left[ a_{\bm{k}}^{(\lambda)} \mcA_\lambda(k,t) 
  + a_{-\bm{k}}^{(\lambda) \dag} \mcA_\lambda^*(k,t) \right]
\,,
\label{quantization}
\end{equation}
where $e^{(\pm)}_i(\hat{\bm{k}})$ are the right/left-handed polarization vectors which satisfy $i\epsilon_{ijl} k_j e_l^{(\pm)}(\hat{\bm{k}})=\pm k e_i^{(\pm)}(\hat{\bm{k}})$,
and 
$a_{\bm{k}}^{(\pm) \dag}, a_{\bm{k}}^{(\pm)}$
are the creation/annihilation operators which satisfy the usual commutation relation, $[a^{(\lambda)}_{\bm{k}},a^{(\sigma) \dag}_{-\bm{k}'}]
= (2\pi)^3\delta(\bm{k}+\bm{k}')\delta^{\lambda \sigma}$.
The equation of motion (EoM) for the mode function of the gauge field $\mcA_\pm$ then is simply the classical equation obtained by varying the action $\int\sqrt{|g|}{\cal L}$ wrt $\mcA_{\pm}$. In a spatially flat Friedmann Universe we find
\begin{equation}
\left[ \partial_\eta^2 +k^2 \pm 2\gamma k \frac{\partial_\eta I}{I}-\frac{\partial_\eta^2I}{I} \right] \Big(I(\eta)\mcA_\pm(k,\eta)\Big)=0\,.
\label{A EoM}
\end{equation}
We use the conformal time $\eta$ as the time variable.
Note that by virtue of the conformal symmetry of the gauge field, 
the above EoM does not depend on the cosmic scale factor $a(\eta)$.

%
\begin{figure}[tbp]
  \begin{center}
  \includegraphics[width=90mm]{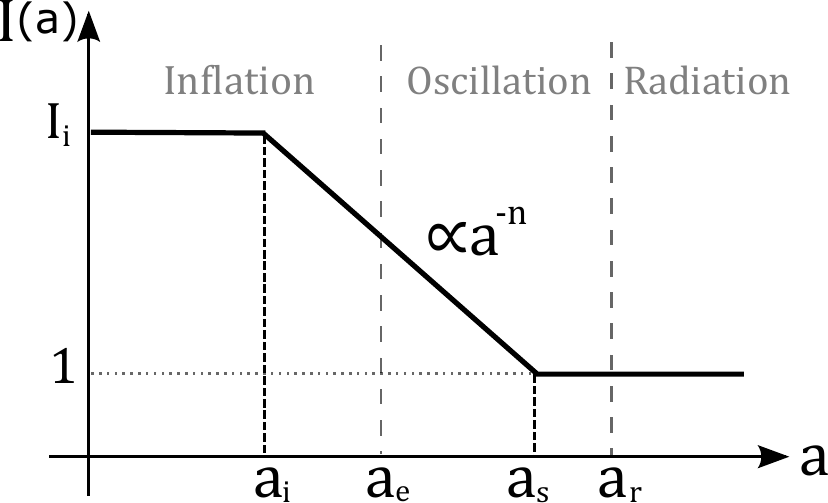}
  \end{center}
  \caption
 {The behavior of $I(a)$ given in eq.~\eqref{simple I}.
 $I(a)$ decreases in proportion to $a^{-n}$ for $a_i<a<a_s$.
 $I$ starts decreasing during inflation $(a_i<a_e)$ and stops
 varying before or the same time as the reheating completion $(a_e<a_s\le a_r)$. We mainly consider $n=3$ which produces scale-invariant magnetic fields.}
 \label{I_behavior}
\end{figure}
%
To solve the above EoM, we need to specify the time evolution of $I(\chi)$.
We assume that $I$ is constant at the beginning, starts varying at a certain time during inflation and stops evolving before or at the completion of reheating. We set $I=1$ when it stops such that the above field is canonically normalized at late time and normal electromagnetism is restored. The behavior of the background cosmic expansion and the kinetic function $I$ are given by
\begin{align}
\eta &= 
\left\{
 \begin{array}{lc}
 -1/a H_\inf \ \propto a^{-1}&  (a <a_e)\\
 2/aH \qquad\propto a^{1/2}\ &  (a_e <a <a_r)
 \end{array} 
 \right.,
\\[3mm]
I(\eta) &= 
\left\{
 \begin{array}{lc}
 (a_i/a_r)^{-n}\equiv I_i & (a <a_i) \\
 (a(\eta)/a_s)^{-n} &  (a_i <a <a_s) \\
 1  & \quad (a_s<a) \\
 \end{array} 
 \right. ,
\label{simple I}
\end{align}
where $a_i, a_e, a_s$ and $a_r$ denote the values of scale factor $a$ when $I(\eta)$ starts varying,  when inflation ends,  when $I$ stops evolving and  when reheating completes, respectively.
Fig.~\ref{I_behavior} illustrates this behavior of $I(a)$.
During reheating we assume a simple matter-dominated expansion, $a\propto \eta^2$.
It should also be stressed that our scenario is free from the strong coupling problem~\cite{Demozzi:2009fu} for $n\ge 0$, because $I$ is never smaller than unity.

\subsection{Solving the dynamics of the electromagnetic field }

Substituting eq.~\eqref{simple I} into \eqref{A EoM}, one finds the EoM for the gauge field mode function reads

\begin{align}
\left[\partial_\eta^2 +k^2\right](I_i \mcA_\pm^{\rm BD})=&0
\,, \qquad (a<a_i )
\label{BD EoM}
\\ 
\left[\partial_\eta^2 +k^2\pm 2\xi \frac{k}{\eta}-\frac{n(n-1)}{\eta^2}\right](I\mcA_\pm^\inf)=&0
\,, \qquad (a_i <a <a_e)
\label{inf EoM}
\\
\left[\partial_\eta^2 +k^2\mp 4\xi \frac{k}{\eta}-\frac{2n(2n+1)}{\eta^2}\right](I\mcA_\pm^\osc)=&0,
\qquad (a_e <a <a_s)
\label{osc EoM}
\\
\left[\partial_\eta^2 +k^2\right] \mcA_\pm^{\rm fin}=&0
\,, \qquad (a_s<a)
\label{RH EoM}
\end{align}
with
\begin{equation}
\xi\equiv n\gamma.
\end{equation}
We solve the EoMs with  Bunch-Davies initial conditions for $a<a_i$,
\begin{equation}
I_i\mcA_\pm^\BD (a<a_i) =\frac{e^{-ik \left( \eta - \eta_i \right)} }{\sqrt{2k}} \; ,
\label{BD vac}
\end{equation}
where the constant phase $e^{ik \eta_i}$ is added for the convenience of calculation.
The solutions for $a>a_i$ can be obtained in the usual way; after finding the general solution of the differential equation, one can fix the integration constants
with the junction condition,
\begin{equation}
\mcA_\pm^{\rm before}(\eta_*) = \mcA_\pm^{\rm after}(\eta_*),
\qquad
\partial_\eta\mcA_\pm^{\rm before}(\eta_*) = \partial_\eta\mcA_\pm^{\rm after}(\eta_*),
\end{equation}
where $\eta_*$ is one of the junction points, $\eta_i, \eta_e$ or $\eta_s$, and $\mcA_\pm^{\rm before}$ and $\mcA_\pm^{\rm after}$ are the mode functions earlier and later than that time, respectively.
We calculate the solutions during inflation and during the inflaton oscillation phase in order.

\subsubsection{During inflation: $a_i<a<a_e$}

The general solution for eq.~\eqref{inf EoM}
 is given by
\begin{equation}
I\mcA_\pm^\inf=\frac{1}{\sqrt{2k}}\left[C_1^\pm\, M_{\mp i\xi,n-\frac{1}{2}}(2i k\eta)+
C_2^\pm\, W_{\mp i\xi,n-\frac{1}{2}}(2ik\eta)\right],
\label{inf gen sol}
\end{equation}
where $M_{\alpha,\beta}(z)$ and $W_{\alpha,\beta}(z)$ are the Whittaker functions~\cite{Abra}
and $C_1^\pm, C_2^\pm$ are integration constants.
Solving the junction condition,
$\mcA_\pm^\BD(\eta_i) = \mcA_\pm^\inf(\eta_i)$ and $\partial_\eta\mcA_\pm^\BD(\eta_i) = \partial_\eta\mcA_\pm^\inf(\eta_i)$,
we obtain%
\footnote{Here we use the identity, $M_{\alpha,\beta}(z) \partial_z W_{\alpha,\beta}(z)-W_{\alpha,\beta}(z) \partial_z M_{\alpha,\beta}(z)=-\Gamma(2\beta+1)/\Gamma(\beta-\alpha+1/2)$.}
\begin{align}
C_1^\pm&=
\frac{\Gamma(n\pm i\xi)}{2ik\eta_i \Gamma(2n)}\left[W_{1\mp i \xi ,n-\frac{1}{2}}(2 i k \eta_i)+(n\mp i\xi-2ik\eta_i ) W_{\mp i \xi ,n-\frac{1}{2}}(2 i k \eta_i)\right],
\\
C_2^\pm&=
\frac{\Gamma(n\pm i\xi)}{2ik\eta_i \Gamma(2n)}\left[(n\mp i\xi ) M_{1\mp i \xi ,n-\frac{1}{2}}(2 i k \eta_i)-(n\mp i\xi-2ik\eta_i ) M_{\mp i \xi ,n-\frac{1}{2}}(2 i k \eta_i)\right].
\label{analytic C2}
\end{align}
In the sub-horizon limit $|k\eta_i|\gg 1$, eq.~\eqref{analytic C2} becomes $C_2^\pm\simeq e^{\pm \frac{1}{2}\pi \xi}$ which reflects the parity violation of the last term in eq.~\eqref{model lagrangian}.
In the super-horizon  limit $|k\eta_i|\ll 1$, however,  
$C_2^\pm \simeq \Gamma(n\pm i\xi)|2k\eta_i|^n/\Gamma(2n)$
suppresses the both polarization modes.

\subsubsection{During the oscillation era: $a_e<a<a_s$}
\label{During the oscillation era: $a_e<a<a_s$}

The general solution for eq.~\eqref{osc EoM} is given by
\begin{equation}
I\mcA_\pm^\osc=\frac{1}{\sqrt{2k}}\left[D_1^\pm\, M_{\pm 2i\xi,2n+\frac{1}{2}}(2i k\eta)+
D_2^\pm\, W_{\pm 2i\xi,2n+\frac{1}{2}}(2ik\eta)\right],
\label{osc gen sol}
\end{equation}
where $D_1^\pm$ and $D_2^\pm$ are integration constants.
The junction condition at the end of inflation $a=a_e$ is
$\mcA_\pm^\inf(\eta_e) = \mcA_\pm^\osc(\tilde{\eta}_e),
\partial_\eta\mcA_\pm^\inf(\eta_e) = \partial_\eta\mcA_\pm^\osc(\tilde{\eta}_e)$.
It should be noted that the conformal time $\eta$ is not continuous here.
Requiring that the scale factor $a$ and Hubble parameter $H$ are continuous,
one finds that the conformal time $\eta$ jumps as
\begin{equation}
\eta_e = -\frac{1}{a_e H_\inf} \quad{\rm (end\ of\ inflation)}
\quad\Longrightarrow\quad
\tilde{\eta}_e= \frac{2}{a_e H_\inf}\quad {\rm (onset\ of\ oscillation)}.
\end{equation}
Solving the above junction condition, we obtain general expressions for $D_1^\pm$ and $D_2^\pm$ which are lengthy.
However, since we are interested only in  modes that exit the horizon during inflation, we need the super-horizon limits (i.e. $ -k\eta_e\ll 1$) of $D_1^\pm$ and $D_2^\pm$,
\begin{align}
D_{1}^\pm &\simeq \frac{2^{-5n}\Gamma(2n)}{(4n+1)\Gamma(n\pm i\xi)}
|k\eta_e|^{-3n}C_2^\pm,
\label{D1 asympt}
\\
D_{2}^\pm &\simeq 3\cdot2^{3n+1}\frac{\Gamma(2n-1)\Gamma(2n+1\mp2i\xi)}{\Gamma(4n+2)\Gamma(n\pm i\xi)}
\left|k\eta_e\right|^{n+1}C_2^\pm.
\end{align}
Here, the both are proportional to $C_2$, because the second term in eq.~\eqref{inf gen sol} is the growing mode while the term proportional to $C_1$ can be neglected at the end of inflation.
On the other hand,  after inflation the first term in eq.~\eqref{osc gen sol} is the growing mode since $|k\eta|$ is now increasing.

\subsubsection{After $I$ stops evolving: $a_s<a$}

The general solution for eq.~\eqref{RH EoM} is  a free electromagnetic wave as in Minkowski spacetime,
\begin{equation}
\mcA_\pm^{\rm fin}=\frac{1}{\sqrt{2k}}\left[F_1^\pm e^{ik(\eta-\eta_s)}+ F_2^\pm e^{-ik(\eta-\eta_s)}\right],
\end{equation}
where $F_1$ and $F_2$ are integration constants. 
The junction condition when $I$ stops  evolving at $a=a_s$ is $\mcA_\pm^\osc(\eta_s) = \mcA_\pm^{\rm fin}(\eta_s),
\partial_\eta\mcA_\pm^\osc(\eta_s) = \partial_\eta \mcA_\pm^{\rm fin}(\eta_s)$.
Solving this junction condition, we obtain
\begin{align}
F_1^\pm &= \frac{-i}{2 k \eta_s}
\Big[ 2 D_1^\pm (ik\eta_s+ n-i\xi ) M_{2 i \xi ,2 n+\frac{1}{2}}+ D_1^\pm (2  n+2i \xi +1) M_{2 i \xi +1,2 n+\frac{1}{2}}
\notag\\
&\qquad\qquad+2 D_2^\pm  (ik\eta_s+ n-i\xi ) W_{2 i \xi ,2 n+\frac{1}{2}}- D_2^\pm W_{2 i \xi +1,2 n+\frac{1}{2}}\Big],
\\
F_2^\pm &= \frac{i}{2 k \eta_s}
\Big[ 2 D_1^\pm ( n-i\xi ) M_{2 i \xi ,2 n+\frac{1}{2}}+ D_1^\pm (2  n+2i \xi +1) M_{2 i \xi +1,2 n+\frac{1}{2}}
\notag\\
&\qquad\qquad+2 D_2^\pm  ( n-i\xi ) W_{2 i \xi ,2 n+\frac{1}{2}}- D_2^\pm W_{2 i \xi +1,2 n+\frac{1}{2}}\Big],
\end{align}
where all the suppressed arguments of the Whittaker functions are $2 i k \eta_s$.
It is interesting to note that the mode function is proportional to
the sin function in the super-horizon limit,
\begin{equation}
\mcA_\pm^{\rm fin}(\eta)\simeq 
\frac{2^{1-n}\Gamma(2n)}{\sqrt{2k}\,\Gamma(n\pm i\xi)}|k\eta_e|^{-n}\left(\frac{a_s}{a_e}\right)^n
C_2^\pm\, \sin(k(\eta-\eta_s)),
\qquad (|k\eta_s|\ll 1)
\label{Afin solution}
\end{equation}
This is because $\mcA_\pm^{\osc}$ was increasing and did not have a constant part which would be inherited by the coefficient of $\cos(k(\eta-\eta_s))$.
For $k\eta_s\ll k\eta\ll 1$, the mode function grows as $\mcA_\pm^{\rm fin}\propto \eta\propto a^{1/2}$ until it reaches the maximum value of the constant oscillation for the first time. 
This slows down the dilution of the magnetic energy density on super-horizon scales from $a^{-4}$ into $a^{-3}$, unless the electric conductivity affects its dynamics. This fact was recently also noticed in Ref.~\cite{Kobayashi:2019uqs}.

\subsection{The electromagnetic spectra}

%
\begin{figure}[tbp]
  \hspace{-2mm}
  \includegraphics[width=70mm]{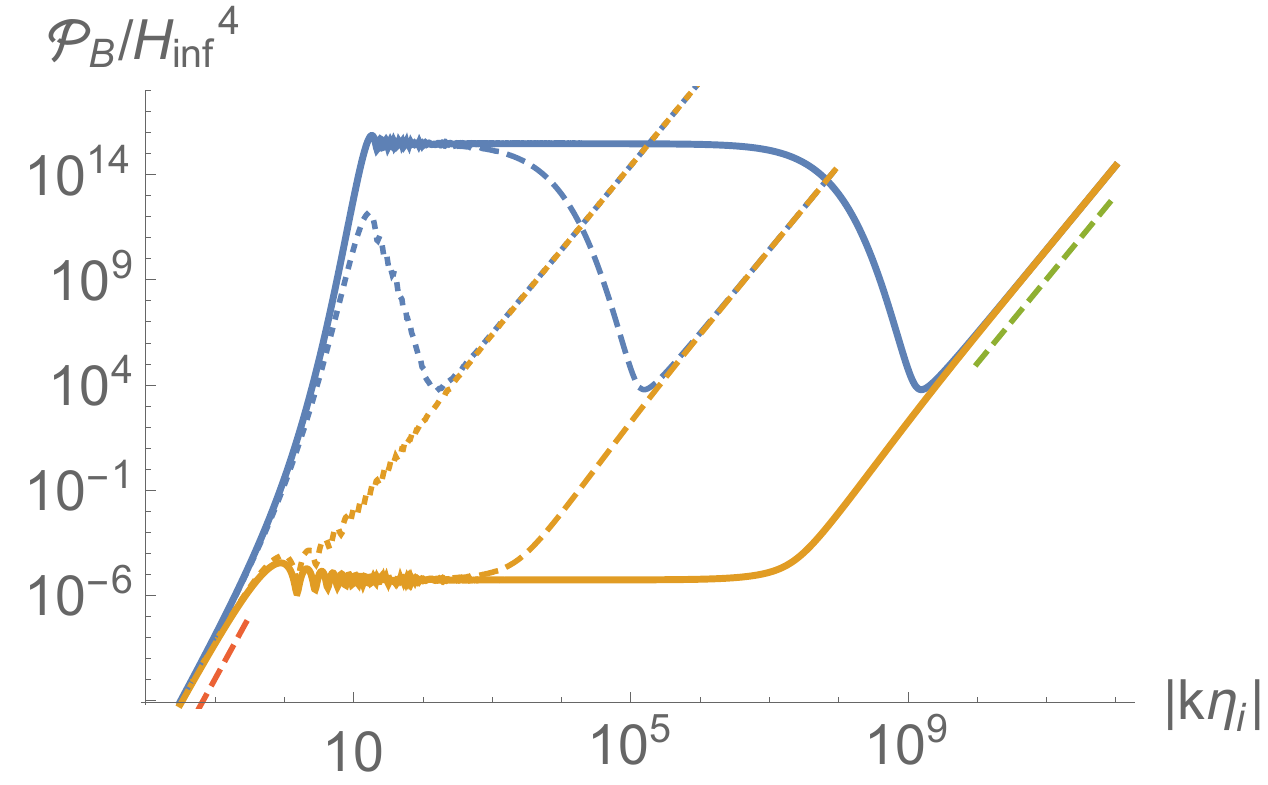}
  \hspace{5mm}
  \includegraphics[width=70mm]{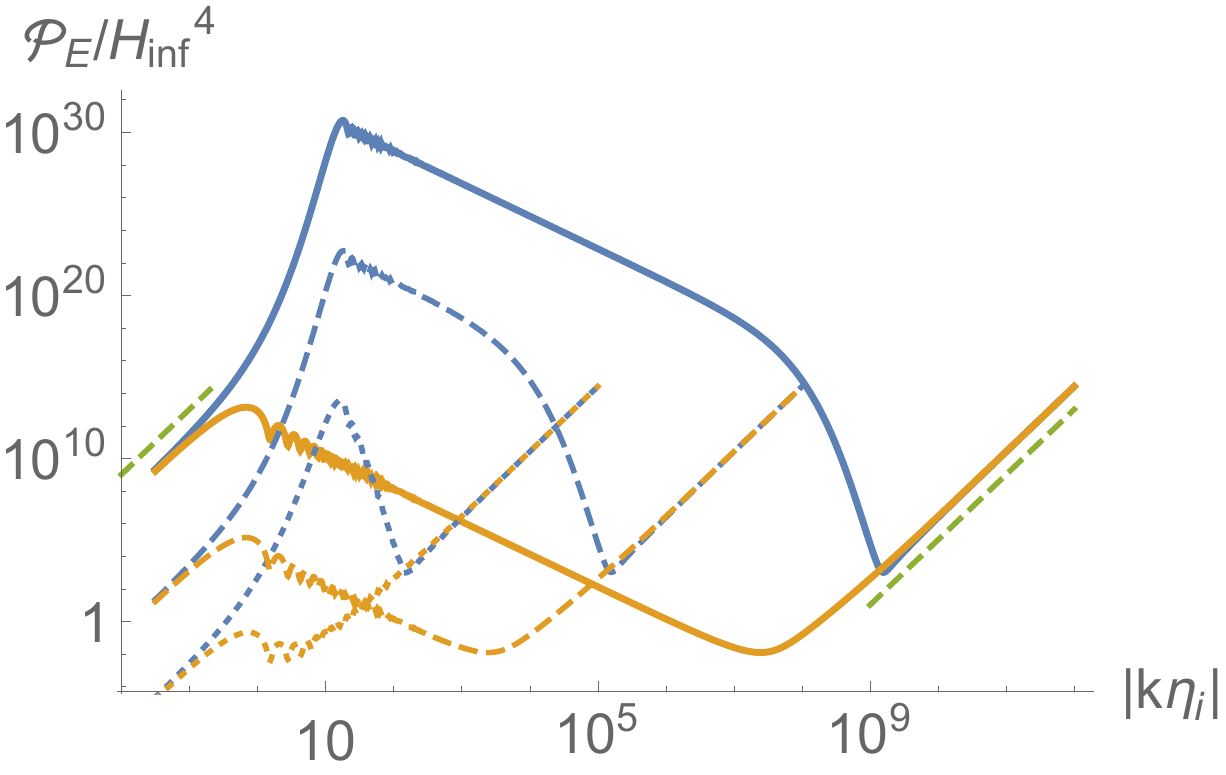}
  \caption
 {$\mcP_B(k,\eta)/H_{\inf}^4$ (left panel) and $\mcP_E(k,\eta)/H_{\inf}^4$ (right panel) of the $+$ (blue) and $-$ (yellow) polarization modes are plotted for $n=3$ and $\xi=7.6$ during inflation. 
The time evolution is shown for $a/a_i=10$ (dotted),
$10^4$ (dashed) and $10^8$ (solid). One can see that only the + mode (blue) is amplified around the horizon crossing due to the tachyonic instability, and both magnetic modes stay constant on super-horizon scales. Hence, a scale-invariant helical magnetic field is produced. However, the modes which have already exited the horizon before $I$ starts varying, $|k\eta_i|\lesssim 1$, are not amplified and have a very blue spectrum $\mcP_B \propto k^6$. The red and green dashed lines show $k^6$ and $k^4$ dependence as references, respectively.}
 \label{PEB inf}
\end{figure}
%
%
\begin{figure}[tbp]
  \hspace{-2mm}
  \includegraphics[width=70mm]{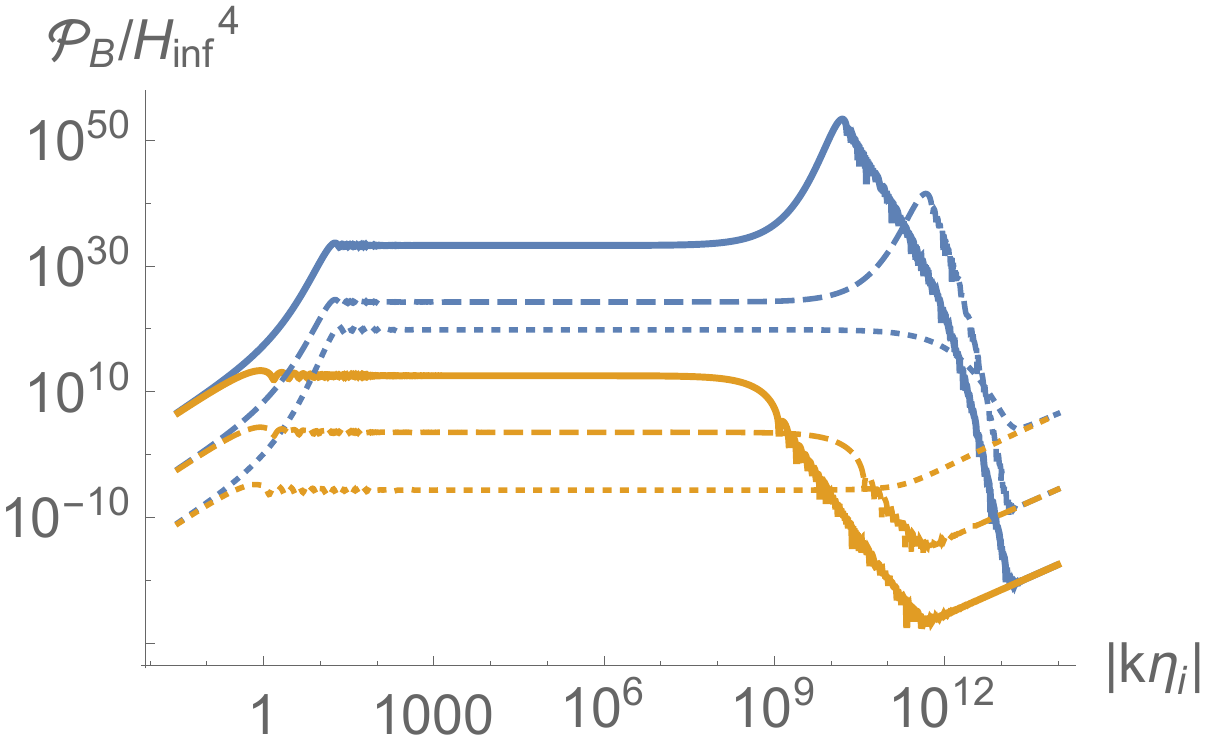}
  \hspace{5mm}
  \includegraphics[width=70mm]{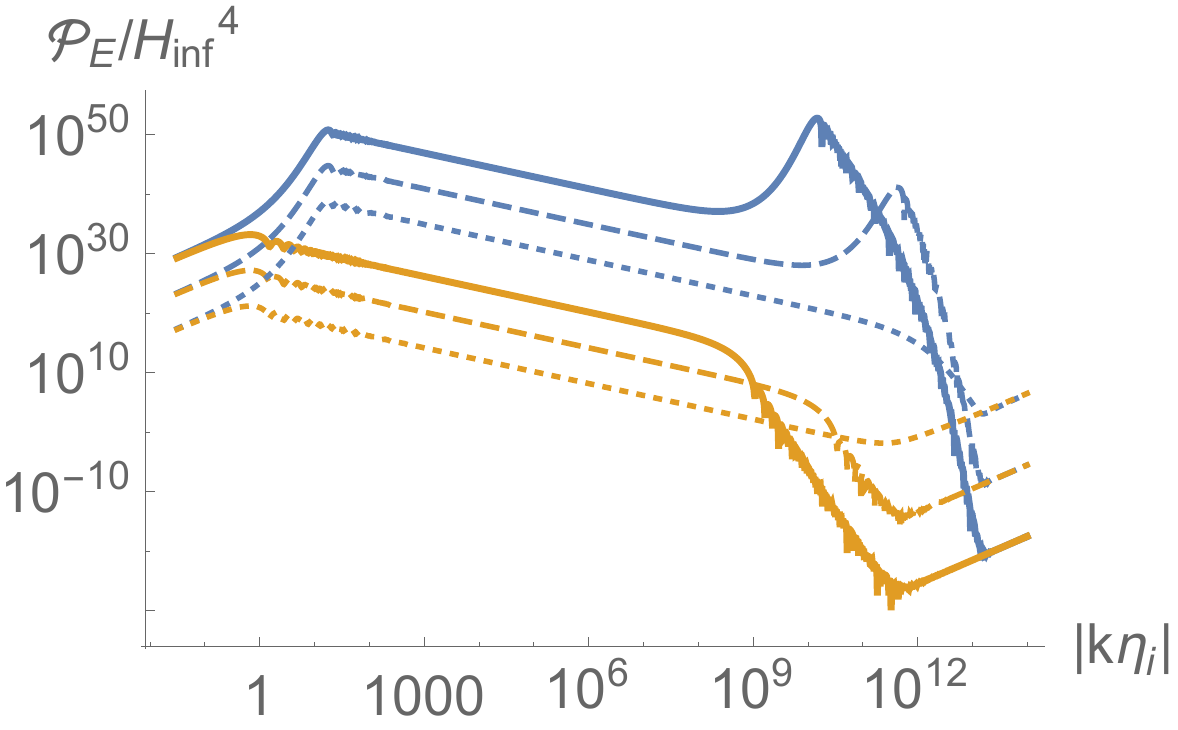}
  \caption
 {The same figure as Fig.~\ref{PEB inf} but during the inflaton oscillation era after inflation. We set $a_e/a_i=10^{12}$ 
and the spectra are  shown at three different times, $a/a_e=1$ (dotted),
$10^3$ (dashed) and $10^{6}$ (solid). 
The super-horizon modes continue  to increase. Furthermore, the $+$ modes are further amplified by the tachyonic instability when they re-enter the horizon. The small wiggles are due to the  oscillations of the modes on sub-horizon scales.
In this figure  electric conductivity which can stop the time evolution of the electromagnetic fields before $a_s$ is not taken into account. For our fiducial parameters, amplification is terminated by conductivity at $a/a_e\simeq 10^4$ as we will see in Sec.~\ref{Magnetic Fields Today}.
 The peak  value of $\mcP_E$ is then much smaller than $\Mpl^2H_{\inf}^2\approx 10^{54}H_{\inf}^4 [(10^{5}\GeV)^{4}/\rho_\inf]$ and backreaction can be safely ignored for sufficiently low inflation scales.
}
 \label{PEB osc}
\end{figure}
%
Having these analytic solutions of $\mcA_\pm(\eta)$, we can compute
the power spectra of the electromagnetic fields and their time evolution.
We introduce the power spectra of the electric and magnetic fields 
for each polarization as
\begin{equation}
\mcP_E^\pm (\eta,k) \equiv \frac{k^3 I^2}{2\pi^2 a^4} |\partial_\eta \mcA_\pm|^2 \; ,\qquad
\mcP_B^\pm (\eta,k) \equiv \frac{k^5 I^2}{2\pi^2 a^4}  |\mcA_\pm|^2 \,.
\label{P of EB}
\end{equation}
Of course, the total power spectra are the sums,
$\mcP_E=\mcP_E^+ +\mcP_E^-,\ \mcP_E=\mcP_B^+ +\mcP_B^-$.
Using the asymptotic behavior of the Whittaker functions,
one can show that the electromagnetic power spectra in the super-horizon limit are given by
\begin{align}
\mcP_B^{\inf}(\eta,k) &\xrightarrow{|k\eta|\ll 1}
\frac{H_\inf^4}{2^{2n}\pi^2} \left|\frac{\Gamma(2n-1)}{\Gamma(n+i\xi)}\right|^2\,
|C_2^\pm|^2  |k\eta|^{6-2n},
\\
\mcP_E^{\inf}(\eta,k) &\xrightarrow{|k\eta|\ll 1}
\frac{H_\inf^4}{2^{2n-2}\pi^2} \left|\frac{\Gamma(2n)}{\Gamma(n+i\xi)}\right|^2\,
|C_2^\pm|^2  |k\eta|^{4-2n},
\\
\mcP_B^{\osc}(\eta,k) &\xrightarrow{|k\eta|\ll 1}
\frac{2^{2-2n}H_\inf^4}{(4n+1)^2\pi^2} \left|\frac{\Gamma(2n)}{\Gamma(n+i\xi)}\right|^2\,
|C_2^\pm|^2  \left(\frac{k}{a_e H_\inf}\right)^{6-2n}
\left(\frac{a}{a_e}\right)^{2n-3},
\label{PB osc etak}
\\
\mcP_E^{\osc}(\eta,k) &\xrightarrow{|k\eta|\ll 1}
\frac{H_\inf^4}{2^{2n}\pi^2} \left|\frac{\Gamma(2n)}{\Gamma(n+i\xi)}\right|^2\,
|C_2^\pm|^2  \left(\frac{k}{a_e H_\inf}\right)^{4-2n}
\left(\frac{a}{a_e}\right)^{2n-4},
\label{PE osc etak}
\\
\mcP_B^{\rm fin}(\eta,k) &\xrightarrow{|k\eta|\ll 1}
\frac{H_\inf^4}{2^{2n-2} \pi^2} \left|\frac{\Gamma(2n)}{\Gamma(n+i\xi)}\right|^2\,
|C_2^\pm|^2  \left(\frac{k}{a_e H_\inf}\right)^{6-2n}
\left(\frac{a_s}{a_e}\right)^{2n-3}\left(\frac{a}{a_s}\right)^{-3},
\label{PB fin etak}
\\
\mcP_E^{\rm fin}(\eta,k) &\xrightarrow{|k\eta|\ll 1}
\frac{H_\inf^4}{2^{2n} \pi^2} \left|\frac{\Gamma(2n)}{\Gamma(n+i\xi)}\right|^2\,
|C_2^\pm|^2  \left(\frac{k}{a_e H_\inf}\right)^{4-2n}
\left(\frac{a_s}{a_e}\right)^{2n-4}\left(\frac{a}{a_s}\right)^{-4},
\end{align}
where $n>1/2$ is assumed and  the asymptotic form of $D_1$ eq.~\eqref{D1 asympt} is used.
The magnetic power spectrum $\mcP_B$ is always proportional to $k^{6-2n}$.
Therefore if we set $n=3$, a scale-invariant magnetic fields will be generated
for $-\eta_i^{-1}\ll k \ll \eta^{-1}$,
\begin{equation}
n=3\quad\Longrightarrow\quad
{\rm Scale\ invarinant\ MF}.
\label{n3 SI}
\end{equation}
In Figs.~\ref{PEB inf}-\ref{PEB osc}, we illustrate the time evolution of the power spectra $\mcP_{E,B}^\pm$ for the case  $n=3$ and $\xi=7.6$ without using super/sub-horizon approximations.
It is confirmed that scale-invariant magnetic fields are generated
and that the above analytic expressions in the super-horizon limit are
good approximations.

\section{Electric Conductivity}
\label{Electric Conductivity}

In the previous section, we have determined the evolution of the electromagnetic field in our model. However, in these calculation they are still electromagnetic waves in which electric and magnetic fields oscillate into each other. 
To connect them to the present cosmic magnetic fields, we need to consider the electric conductivity which induces a conversion of  electromagnetic waves into frozen magnetic fields and completely damps the electric field. 
In this section, we introduce conductivity and solve the dynamics of the electromagnetic fields in a highly conducting plasma as the generated during reheating.

\subsection{The effect of conductivity on magnetogenesis}

We consider the interaction between the charge current and the electromagnetic field in our model lagrangian eq.~\eqref{model lagrangian}, 
\begin{equation}
\mathcal{L}_{\rm int}=-J^\mu A_\mu,
\label{L int}
\end{equation}
where the current is phenomenologically given by~\cite{Martin:2007ue}%
\footnote{We neglect $\sigma_c (\bm v\times \bm B)$ in the r.h.s. of eq.~\eqref{current eq} as a higher order term in the cosmological perturbation. Note that both $\bm v$ and $\bm B$ are perturbations in the Friedman universe and therefore small.}
\begin{equation}
J_\mu =\rho_e u_\mu +\sigma_c E_\mu,
\label{current eq}
\end{equation}
where $\rho_e$ is the charge density which is assumed to be negligible
for the neutrality of the universe on the scale in interest.
Here the electric conductivity $\sigma_c$ is introduced as a proportionality constant between the current density and the electric field, $\bm J=\sigma_c \bm E$ (see Appendix.~\ref{Derivation of the conductivity} for a derivation).
In Coulomb gauge, this new term in the Lagrangian Eq.~\eqref{L int} modifies the EoM for photon, Eq.~\eqref{A EoM}, into 
\begin{equation}
\partial_\eta^{2}\mcA_\pm+\left[2\frac{\partial_\eta I}{I}+\frac{a\sigma_c}{I^2}\right]\partial_\eta\mcA_\pm
+\left[k^2\pm 2\gamma k \frac{\partial_\eta I}{I}\right]\mcA_\pm=0,
\label{EoM w/ sigma}
\end{equation}
where we have used $E_i=-\dot{A}_i=-a^{-1}\partial_\eta A_i$.
 Changing the time variable to  cosmic time, $dt=ad\eta$, the above EoM can be written as
\begin{equation}
\ddot{\mcA}_\pm+\left[H+\frac{\sigma_c}{I^2}+2\frac{\dot{I}}{I}\right]\dot{\mcA}_\pm
+\left[\frac{k^2}{a^2}\pm2\gamma \frac{k}{a}\frac{\dot{I}}{I}\right]\mcA_\pm=0.
\end{equation}
In this expression one  sees that $\sigma_c>0$ works as a friction term, while $\dot{I}/I=-nH<0$ is amplifying the electromagnetic fields and can be interpreted as a ``negative friction''. Note that since $\sigma_c$ is divided by $I^2$, the conductivity term is suppressed for $I\gg 1$. This is because, in the weak coupling regime ($I\gg 1$),  electric conductivity which is caused by the coupling to the charged current becomes inefficient. 

To understand how the electric conductivity affects the generation of the electromagnetic fields, we analyze a simple but sufficiently general case which is analytically solvable.
In addition to the assumption $I=(\eta/\eta_s)^{-2n}$
for $\eta_e<\eta<\eta_s$ and $I=1$ for $\eta_s<\eta$ (see eq.~\eqref{simple I}), we also assume power-law time dependence for the conductivity after inflation,
\begin{equation}
\sigma_c(\eta) = \sigma_r\left(\frac{a}{a_r}\right)^m=\sigma_r\left(\frac{\eta}{\eta_r}\right)^{2m}
\qquad (\eta_e<\eta\le\eta_r)\,,
\label{m introduced}
\end{equation}
where $m$ is a constant and $\sigma_r$ is the conductivity at the reheating completion.
With this ansatz, the conductivity term in Eq.~\eqref{EoM w/ sigma} becomes
\begin{equation}
\frac{a\sigma_c}{I^2}=\frac{2\sigma_r}{H_r \eta_r}\left(\frac{\eta}{\eta_r}\right)^{2m+2}\times
\begin{cases}
(\eta/\eta_s)^{4n} & (\eta_e<\eta<\eta_s) \\
1 & (\eta_s<\eta\le\eta_r) \,. \\
\end{cases}
\end{equation}

After the end of inflation and before $I$ stops varying, $\eta_e<\eta<\eta_s$, Eq.~\eqref{EoM w/ sigma} reads
\begin{equation}
\partial_\eta^{2}\mcA_\pm+\frac{4n}{\eta}\left[-1+\left(\frac{\eta}{\bar{\eta}_c}\right)^{2m+4n+3}\right]\partial_\eta\mcA_\pm
+\left[k^2\mp \gamma k \frac{4n}{\eta}\right]\mcA_\pm=0,
\label{EoM compare sigma}
\end{equation}
where the conductivity term becomes comparable to the negative friction term at $\bar{\eta}_c$,
\begin{equation}
\bar{\eta}_c=\left[2n\frac{H_r}{\sigma_r}\,\eta_r^{2m+3}\eta_s^{4n}\right]^{\frac{1}{2m+4n+3}}.
\label{etac def}
\end{equation}
Note if this expression results  in $\bar{\eta}_c\gg\eta_s$,
 conductivity is always negligible while $I$ is varying.

%
\begin{figure}[tbp]
  \hspace{-2mm}
  \includegraphics[width=70mm]{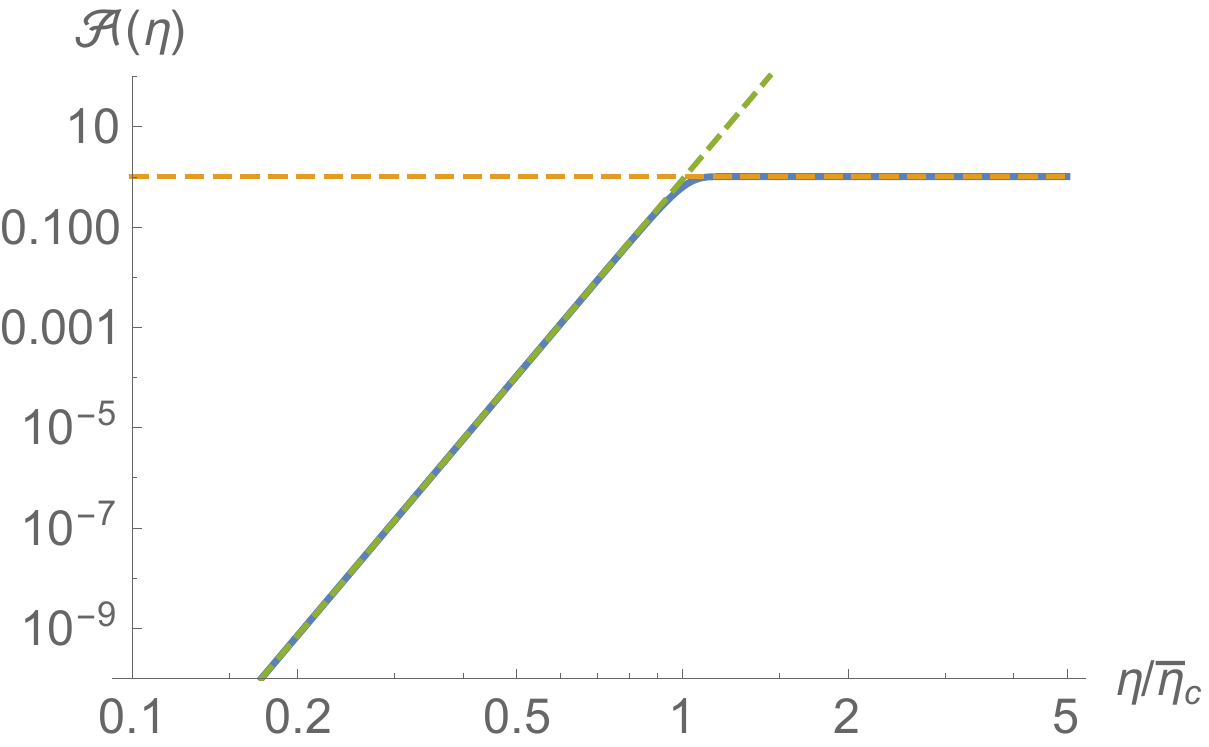}
  \hspace{5mm}
  \includegraphics[width=70mm]{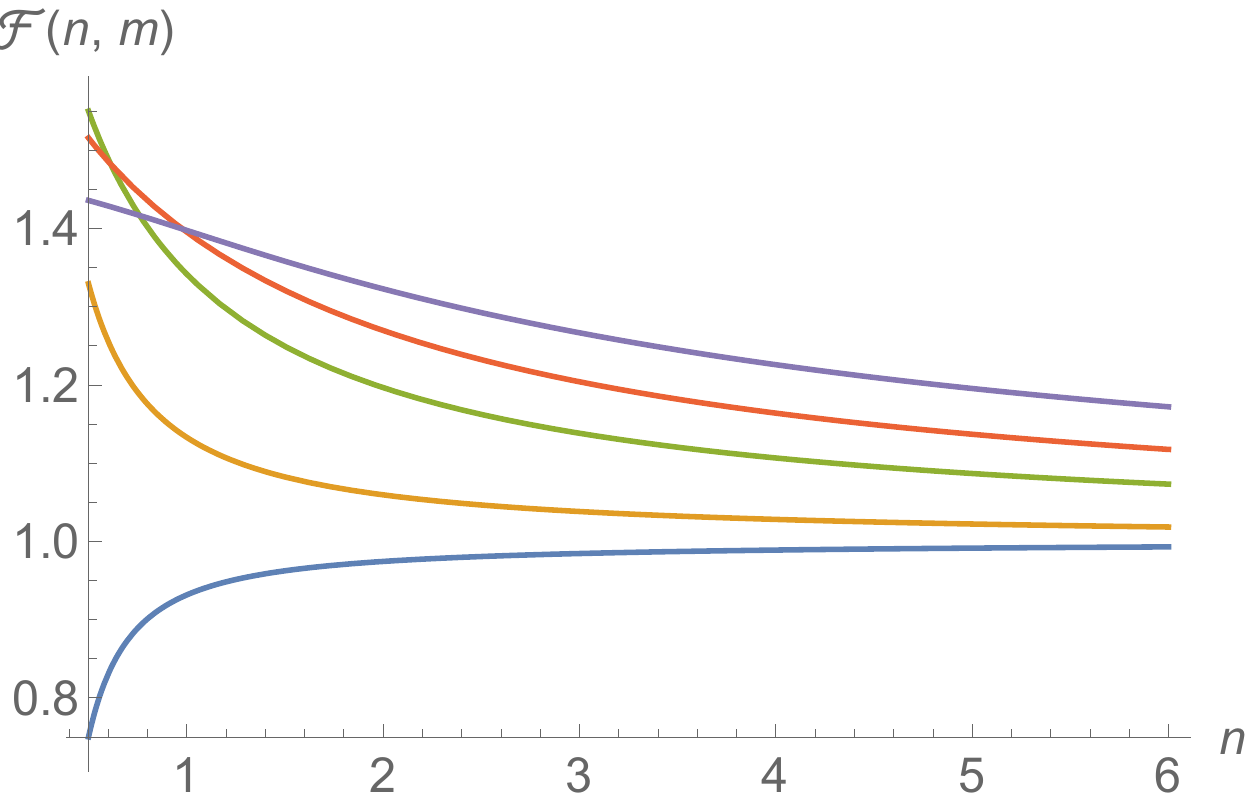}
  \caption
  {{\bf (Left panel)} The analytic solution eq.~\eqref{sol w/ sigma} is shown as blue solid line for $G_1=1, n=3$ and $\ell=15$. The asymptotic forms, $\mcA(\eta\ll\eta_c)=4n (\eta/\eta_c)^{4n+1}/(\ell\Gamma(\frac{4n+\ell+1}{\ell}))$ and $\mcA(\eta\gg\eta_c)=1$, are also plotted as dashed lines.
One can see that the growth of $\mcA(\eta)$ is terminated by the electric conductivity at $\eta\simeq \bar{\eta}_c$.
 {\bf (Right panel)} $\mathcal{F}(n,m)$ introduced in Eq.~\eqref{sigma connection}.
 The horizontal axis denotes $n$ and the colored lines represent $m=-2$ (blue), $-3/2$ (orange),
 $-1/4$ (green), $1$ (red) and $3$ (purple) from bottom to top.
 $\mathcal{F}(n,m)\simeq 1$ holds in broad parameter space.}
 \label{Fnm_contour}
\end{figure}
%
Focusing on super-horizon modes and ignoring the third term, 
eq.~\eqref{EoM compare sigma} can be analytically solved as
\begin{equation}
\mcA(\eta)=G_1+G_2\,\Gamma \left(\frac{4 n+1}{\ell},\,
\frac{4n}{\ell}\,
\left(\frac{\eta}{\bar{\eta}_c}\right)^{\ell}\right),
\qquad (\eta_e<\eta<\eta_s)
\label{sol w/ sigma}
\end{equation}
where $\ell\equiv 2 m+4 n+3$, $G_1$ and $G_2$ are integration constants, and $\Gamma(a,x)$ is the incomplete Gamma function.
Here we drop the label of the circular polarization $\lambda=\pm$.
This solution is shown in the left panel of  Fig.~\ref{Fnm_contour}.
Using the asymptotic behavior of the incomplete Gamma function,
\begin{equation}
\Gamma(a,x) =\begin{cases}\Gamma(a)-x^{a}/a & (x\to 0) \\
x^{a-1}\exp[-x] & (x\to \infty) \\
\end{cases},
\end{equation}
one finds that the asymptotic behavior of Eq.~\eqref{sol w/ sigma} is
\begin{equation}
\mcA(\eta) =\begin{cases}G_1+G_2\Gamma(\frac{4 n+1}{\ell})-G_2\frac{\ell}{4n+1}\left(\frac{4n}{\ell}\right)^{\frac{4n+1}{\ell}}\, \left(\frac{\eta}{\bar{\eta}_c}\right)^{4n+1} & (\eta\ll \bar{\eta}_c) \\
G_1+G_2\, \left(\frac{4n}{\ell}\right)^{\frac{4n+1}{\ell}-1}\left(\frac{\eta}{\bar{\eta}_c}\right)^{4n+1-\ell}\exp\left[-\frac{4n}{\ell}\,
\left(\frac{\eta}{\bar{\eta}_c}\right)^{\ell}\right] & (\eta\gg \bar{\eta}_c) \\
\end{cases}.
\end{equation}
The integration constants $G_1$ and $G_2$ are constrained to $G_1+G_2\Gamma(\frac{4 n+1}{\ell})=0$ by requiring the mode function to reproduce the behavior without conductivity
(i.e. $\mcA\sim \eta^{4n+1}$ which was studied in Sec.~\ref{During the oscillation era: $a_e<a<a_s$}) at early times, $\eta\ll \bar{\eta}_c$
where conductivity is negligible.
On the other hand, the mode function quickly tends to a constant $G_1$ once conductivity becomes effective, for $\eta\gg \bar{\eta}_c$. 
Therefore, parameterizing the coefficient of the early-time behavior $\mcA\propto (\eta/\bar{\eta}_c)^{4n+1}$ as $\mcA_c$, the late-time amplitude is simply $\mcA_c$ with an $\mathcal{O}(1)$ correction,
\begin{equation}
\mcA(\eta\ll \bar{\eta}_c<\eta_s)\simeq \mcA_c \left(\frac{\eta}{\bar{\eta}_c}\right)^{4n+1}
\quad\Longrightarrow\quad 
\mcA(\bar{\eta}_c\ll\eta<\eta_s)=\mathcal{F}(n,m)\,\mcA_c,
\label{sigma connection}
\end{equation}
where the correction factor $\mathcal{F}(n,m)\equiv\left(\frac{\ell}{4n}\right)^{\frac{4n+1}{\ell}}
\Gamma\left(1+\frac{4n+1}{\ell}\right)$ is always around unity as shown in Fig.~\ref{Fnm_contour}.
The frozen amplitude of $\mcA$ takes almost the same value as the case where the behavior without the conductivity $\mcA\propto \eta^{4n+1}$
continues up to $\eta=\bar{\eta}_c$ and $\mcA$ suddenly stops there. 
The condition that the conductivity becomes significant before $I$ stops varying is
\begin{equation}
\bar{\eta}_c <\eta_s
\quad\Longrightarrow\quad
\frac{\sigma_r}{H_r}> 2n\left(\frac{\eta_r}{\eta_s}\right)^{2m+3}.
\end{equation}

Note that the analysis so far from eq.~\eqref{EoM compare sigma} have considered
the dynamics after the inflation end and before $I$ stops running, $\eta_e<\eta<\eta_s$. However, even later, when $I\equiv 1$, a very similar argument holds.
By taking the limit $n\to 0$ in eq.~\eqref{EoM compare sigma}, we obtain an analytic solution after $I$ stops evolving,
\begin{equation}
\mcA(\eta)=G_1+G_2\,\Gamma \left(\frac{1}{2m+3},\,
\frac{2}{2m+3}\,
\left(\frac{\eta}{\tilde{\eta}_c}\right)^{2m+3}\right),
\qquad (\eta_s<\eta<\eta_r).
\label{sol w/ sigma2}
\end{equation}
with
\begin{equation}
\tilde{\eta}_c = \left(H_r/\sigma_r\right)^{\frac{1}{(2m+3)}}\eta_r.
\label{etact def}
\end{equation} 
As early time limit $\eta/\tilde{\eta}_r\to0$, eq.~\eqref{sol w/ sigma2} has a constant mode and a growing mode which is proportional to $\eta$. As discussed below eq.~\eqref{Afin solution}, the super-horizon mode obeys $\mcA^{\rm fin}\propto \eta$ in this regime without the conductivity. Hence, the same argument as eq.~\eqref{sigma connection} yields in this case
\begin{equation}
\mcA(\eta_s<\eta\ll \tilde{\eta}_c)\simeq \tilde{\mcA}_c \frac{\eta}{\tilde{\eta}_c}
\quad\Longrightarrow\quad 
\mcA(\eta_s<\tilde{\eta}_c\ll \eta)=\tilde{\mathcal{F}}(m)\,\tilde{\mcA}_c,
\end{equation}
where the correction factor now is $\tilde{\mathcal{F}}(m)=2^{\frac{-1}{2m+3}}(2m+3)^{-\frac{2m+2}{2m+3}}\Gamma(\frac{1}{2m+3})$
which is always close to unity for $m\gtrsim-1$.
Therefore, in a similar way to the case with $\bar{\eta}_c<\eta_s$,
the electric conductivity freezes the magnetic fields on super horizon scales when it becomes significant at $\eta=\tilde{\eta}_c>\eta_s$.

\subsection{When does conductivity terminate magnetogenesis?}
\label{When the conductivity stops magnetogenesis}

In the previous subsection, we solved the dynamics of the electromagnetic field under the effect of the electric conductivity and found that it terminates the growth of the mode function $\mcA_\pm$.
To calculate the final amplitude of the magnetic fields, therefore, it is crucial to estimate when the conductivity becomes significant.
We have given  expressions for this time in terms of, $\bar{\eta}_c$ and $\tilde{\eta}_c$ in eqs.~\eqref{etac def} and \eqref{etact def}, depending on whether conductivity becomes efficient during magnetogenesis or only after, we now define $\eta_c$ as the unified variable
\begin{equation}
\eta_c\equiv \begin{cases}\left[2n\frac{H_r}{\sigma_r}\,\left(\frac{\eta_r}{\eta_s}\right)^{2m+3}\right]^{1/\ell}\eta_s & \qquad\left(\sigma_r/H_r\gg\left(\eta_r/\eta_s\right)^{2m+3} \right) \\
\left(\frac{H_r}{\sigma_r}\right)^{\frac{1}{(2m+3)}}\eta_r & 
\qquad\left(\sigma_r/H_r \lesssim \left(\eta_r/\eta_s\right)^{2m+3} \right)\,. \\
\end{cases}
\label{define etac}
\end{equation}
In this equation, two dimensionless parameters, $\sigma_r/H_r$ and $m$ are involved.
To develop our understanding of $\eta_c$, we now determine these parameters. 

We first compute $\sigma_r/H_r$. It has been known that the conductivity from charged particles in thermal equilibrium is given by~\cite{Baym:1997gq}
\begin{equation}
\sigma_c= c_\sigma T,
\qquad
c_\sigma \simeq \alpha_e^{-1}\approx 10^2,
\label{sigma100T}
\end{equation}
where $\alpha_e$ is the fine structure constant.
Once reheating is completed, $\eta=\eta_r$, the universe is dominated
by radiation and its energy density is $\rho_{\rm rad}=3\Mpl^2 H_r^2 = \pi^2g_*T_r^4/30$, where $T_r$ is the reheating temperature. 
Hence, we find that the conductivity at $\eta_r$ is given by
\begin{equation}
\frac{\sigma_r}{H_r}=c_\sigma\left(\frac{90\Mpl^2}{\pi^2g_* H_r^2}\right)^{\frac{1}{4}}
\approx 10^{20}\,\left(\frac{T_r}{1\GeV}\right)^{-1}\left(\frac{c_\sigma}{10^2}\right)
\left(\frac{g_*}{10^2}\right)^{-\frac{1}{2}}.
\end{equation}
The electric conductivity largely dominates the Hubble scale at the end of reheating.
Note that $\sigma_r/H_r\gg 1$ guarantees that the conductivity becomes significant before the end of reheating for $m>-3/2$ (see eq.~\eqref{etact def}).
Incidentally, since $T\propto a^{-1}$ and $H\propto a^{-2}$ in the radiation dominant era, the ratio $\sigma_c/H$ even increases after reheating.

Let us specify $m$ which parameterizes the time-dependence of $\sigma_c$ as given in Eq.~\eqref{m introduced}. To calculate the temperature during reheating, we need to model how charged particles are produced. 
As an simple model, we consider that  the inflaton decays into charged particles with  constant decay rate $\Gamma_\phi$. 
The energy density of the charged particles $\rho_{\rm cp}$ for $\eta_e\ll \eta\ll \eta_r$ is then given by~\cite{Kolb:1990vq}
\begin{equation}
\rho_{\rm cp}\simeq \frac{6}{5}\Mpl^2 \Gamma_\phi H \propto a^{-3/2}
\quad\Longrightarrow\quad
\sigma_c\propto T\propto a^{-3/8}
\quad\Longrightarrow\quad
m=-\frac{3}{8},
\label{rhorad for const Gphi}
\end{equation}
where we used $\rho_{\rm cp}\propto T^4$.
Note that we assumed that the decay products have  reached thermal equilibrium, otherwise their temperature $T$ is ill-defined. 
We also discuss the out-of-equilibrium case in Appendix.~\ref{Derivation of the conductivity}. 

Let us now come back to eq.~\eqref{define etac} for the present case.
Setting $m=-3/8$, the boundary of the condition between the two cases of eq.~\eqref{define etac} is 
\begin{equation}
\frac{\sigma_r}{H_r}\simeq \left(\frac{\eta_r}{\eta_s}\right)^{2m+3}
\quad\Longrightarrow\quad
\frac{a_r}{a_s} \simeq 5\times 10^{17}
\left(\frac{T_r}{1\GeV}\right)^{-\frac{8}{9}}\left(\frac{c_\sigma}{10^2}\right)^{\frac{8}{9}}
\left(\frac{g_*}{10^2}\right)^{-\frac{4}{9}}.
\label{aras relation}
\end{equation}
%
%
\begin{figure}[tbp]
  \begin{center}
  \includegraphics[width=90mm]{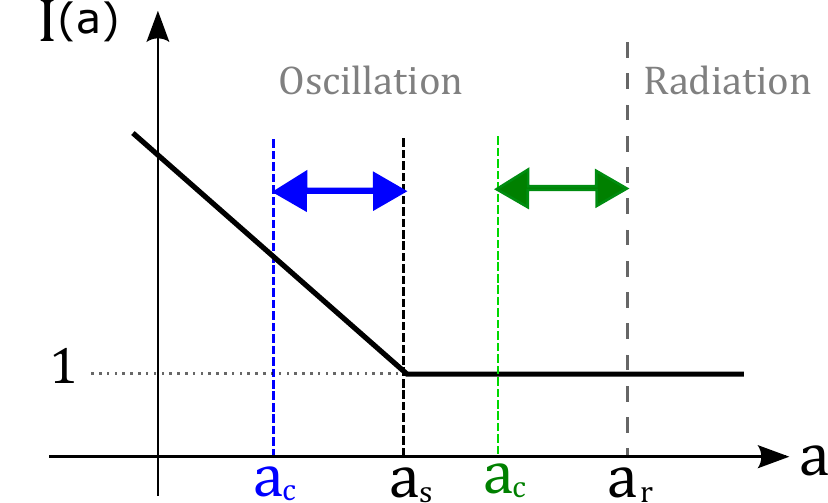}
  \end{center}
  \caption
 {The two possibilities of the timing when the conductivity becomes important $a_c$ derived in eq.~\eqref{ac result}. The first and second line of eq.~\eqref{ac result} corresponds to the blue and green case, respectively.}
 \label{a_conductivity}
\end{figure}
%
Changing $\eta_c$ into the scale factor, $a_c=a(\eta_c)$, one can rewrite eq.~\eqref{define etac} as
\begin{equation}
a_c\equiv \begin{cases}2\times10^{-3}T_{\rm 1GeV}^{\frac{8}{57}}\, (a_r/a_s)^{\frac{3}{19}}\,a_s
& \quad\left(a_r/a_s\ll 10^{18} T_{\rm 1GeV}^{-\frac{8}{9}} \right) \\
2\times 10^{-18} T_{\rm 1GeV}^{\frac{8}{9}}\,a_r & 
\quad\left(a_r/a_s\gtrsim 10^{18} T_{\rm 1GeV}^{-\frac{8}{9}} \right) \\
\end{cases},
\label{ac result}
\end{equation}
where $T_{\rm 1GeV}\equiv T_r/(1\GeV)$ and the dependences on $c_\sigma$ and $g_*$ are suppressed.
This result is illustrated in fig.~\ref{a_conductivity}.
The above equation implies that the electric conductivity terminates the growth of the magnetic fields slightly before the kinetic function $I$ stops, $1>a_c/a_s>0.002T_{\rm 1GeV}^{8/57}$,
unless we have a sufficiently long time interval between the end of magnetogenesis  and the end of reheating. 
This is due to the large kinetic function $I\gg 1$ which substantially suppresses  conductivity until $I$ becomes $\mathcal{O}(1)$ at  $\eta \simeq \eta_s$.
If the time interval between $a_s$ and $a_r$ is long enough, on the other hand, the time interval between the termination of magnetogenesis and the end of reheating is given by $a_r/a_c \simeq 5\times 10^{17} T_{\rm 1GeV}^{-8/9}$, irrespective of $a_s$. 
This is simply because $a_r/a_c$ is determined by the time evolution of $\sigma_c$ during reheating which is fully determined by $\sigma_c/H_c$ and $m$ in this case.

\section{Magnetic Fields Today}
\label{Magnetic Fields Today}

In this section, we calculate the present strength of the magnetic fields produced in our model.

\subsection{Late-time evolution}

Even though our magnetic fields are fully helical, an inverse cascade is not effective in the case of a (nearly)  scale invariant spectrum. This has been shown by numerical simulations in~\cite{Brandenburg:2017rcb,Kahniashvili:2016bkp} and it is not very surprising. The inverse cascade is actually a consequence of helicity conservation. When small scale magnetic fields are damped by viscosity and Alvf\'en wave damping, their helicity has to move to larger scales. If the power is anyway dominated by large scale fields this is a very small effect and will not change the magnetic field spectrum on large scales significantly. Here we simply neglect this small effect.

Therefore, the magnetic power spectrum decays as $\mcP_B\propto a^{-4}$ after the mode function is frozen out, $\mcA_\pm=$const. due to electric conductivity at $\eta=\eta_c$.
Setting $n=3$ and multiplying $\mcP_B(k,\eta_c)$ by $a_c^4$, we obtain 
the present strength of the scale-invariant part of the produced magnetic fields as
\begin{equation}
\mcP_B^{\rm now}(k_i< k< k_T,n=3) \simeq
\frac{H_\inf^4}{2^{4} \pi^2}
\frac{e^{\pi\xi}\,\Gamma^2(6)}{|\Gamma(3+i\xi)|^2}\,
\frac{1}{a_e^3}\times
\begin{cases}a_c^7 & (a_c<a_s) \\a_s^6 a_c  & (a_s<a_c) \\\end{cases},
\end{equation}
where $\mcP^{\rm osc}_B(k,\eta_c)$ in eq.\eqref{PB osc etak}
or $\mcP^{\rm fin}_B(k,\eta_c)$ in eq.\eqref{PB fin etak} are used depending on whether $\eta_c \gtrless\ \eta_s$ and we approximate the correction factors $\mathcal{F}(m,n)$
and $\tilde{\mathcal{F}}(m)$ by unity. 
Here $k_i$ and $k_T$ denote
the IR cut-off scale $k_i= -1/\eta_i$ and the turbulence scale today, respectively.
Inserting numbers, one can rewrite the above equation as
\begin{equation}
\mcP_B^{\rm now}\approx 3\times 10^{-28}\G^2\ \frac{e^{\pi\xi}\sinh(\pi\xi)}{4\xi+5\xi^3+\xi^5}
\left(\frac{\rho_\inf^{1/4}}{10^{5}\GeV}\right)^{12}\left(\frac{T_r}{1\GeV}\right)^{-8}
\left(\frac{g_\ast}{10^2}\right)^{-\frac{7}{3}}
\times
\begin{cases}(a_c/a_r)^7 & (a_c<a_s) \\
a_s^6 a_c/a_r^7  & (a_s<a_c) \\\end{cases},
\label{PBnow1}
\end{equation}
where we evaluated $a_e$ and $a_r$ as
\begin{align}
\ln \left(\frac{a_r}{a_e}\right) 
&\approx 22.8+\frac{4}{3} \ln \left(\frac{\rho_\inf^{1/4}}{10^{5}\GeV}\right)
-\frac{4}{3}\ln \left(\frac{T_r}{1\GeV}\right)-\frac{1}{3}\ln \left(\frac{g_*}{100}\right),
\\
a_r & 
\approx 8 \times10^{-14} \left(\frac{T_r}{1\GeV}\right)^{-1}\left(\frac{g_\ast}{100}\right)^{-1/3}.
\end{align}
Here, the unit conversion, $1\GeV=3.8\times 10^9 \G^{1/2}$, is  used.

\subsection{Consistency conditions}

For successful primordial magnetogenesis, one must satisfy the following three conditions:
The scenario is free from (i) a strong coupling regime for which calculations cannot be trusted, (ii) significant back reaction from the electromagnetic fields which alters the background evolution of the universe (iii) inconsistency with  CMB observations which particularly fixes the power spectrum of curvature perturbations as $\mcP_\zeta (k_{\rm CMB})\approx 2.2\times 10^{-9}$ with $k_{\rm CMB}=0.05\Mpc^{-1}$.
Since the condition (i) is trivially satisfied in our case where $I\ge 1$, we address the backreaction problem (ii) and the curvature perturbation condition (iii).

To satisfy the backreaction condition, the energy fraction of the produced electromagnetic fields should always be a very small fraction of the energy density of the Universe,
\begin{equation}
\Omega_{\em}\equiv \frac{\rho_{\em}}{\rho_{\tot}}\ll1,
\qquad
\rho_\em=\frac{1}{2}\int \frac{\dd k}{k}\Big(\mcP_E(k)+\mcP_B(k)\Big),
\end{equation}
where $\rho_\tot=3\Mpl^2H^2$ denotes the total energy of the universe.
$\Omega_{\em}(\eta)$ during the inflaton oscillation phase is written as
\begin{equation}
\Omega_{\em}(\eta_e<\eta<\eta_r)\simeq \frac{1}{2\rho_\tot}\int\frac{\dd k}{k}\mcP_E^+(k_i,\eta)
=e^{2N_i}\frac{150H_{\inf}^2 a^5}{\pi^3 \Mpl^2 a_e^5}\frac{\mathcal{I(\xi)} \sinh(\pi\xi)}{4\xi+5\xi^3+\xi^5},
\end{equation}
where $N_{i}\equiv \ln(\eta_i/\eta_e)$, we ignore the sub-leading contributions from $\mcP_E^-$ and $\mcP_B^\pm$ and
use eq.~\eqref{PE osc etak} with $n=3$.
A numerical evaluation of $\mathcal{I}(\xi)$ which contains the $k$-integral shows that it is well approximated by (see Appendix~\ref{Numerical Calculation of} for detail)
\begin{equation}
\mathcal{I}(\xi)\equiv \int_0^\infty \frac{\dd y}{y^3}\, \left|C^+_2(y)\right|^2
= \frac{e^{\pi \xi}}{\alpha \xi^2+\beta\xi+\gamma},
\qquad
\alpha\approx 4.7,\ \beta\approx2.7,\ \gamma\approx10,
\label{I integrand}
\end{equation}
where the dummy variable $y\equiv -k\eta_i$ is introduced and the numerical fit is performed for $1\le \xi \le 50$.
$\Omega_\em$ reaches its maximum value at either $\eta_c$ or $\eta_s$ whichever is earlier. We obtain its maximum value given by
\begin{eqnarray}
\Omega_\em^{\max} &\simeq& 10^{-2}\frac{\mathcal{I(\xi)} \sinh(\pi\xi)}{4\xi+5\xi^3+\xi^5}
\left(\frac{\rho_\inf^{1/4}}{10^{5}\GeV}\right)^{12}\left(\frac{T_r}{1\GeV}\right)^{-6}
\left(\frac{g_\ast}{10^2}\right)^{-\frac{5}{3}}\left(\frac{k_i}{1\Mpc^{-1}}\right)^{-2} \nonumber \\
&& \times
\begin{cases}(a_c/a_r)^5 & (a_c<a_s) \\(a_s/a_r)^5 & (a_s<a_c) \,, \\\end{cases}
\end{eqnarray}
where we used $N_i\approx 28+\frac{2}{3}\ln(\rho_{\inf}^{1/4}/10^{5}\GeV)
+\frac{1}{3}\ln(T_r/1\GeV)-\ln(k_i/1\Mpc^{-1})$.  
A lower inflationary energy scale and higher reheating temperature  decrease $\Omega_\em^{\max}$, because they shorten the inflaton oscillation phase in which $\Omega_\em\propto a^5$ grows rapidly.
$\Omega_\em^{\max}$ also depends on the IR cut-off as $\Omega_\em^{\max}\propto k_i^{-2}$,
because the electric power spectrum is red-tilted $\mcP_E\propto k^{-2}$
for $n=3$ and the dominant contribution to $\Omega_\em$ comes from the largest amplified scale $k\simeq k_i$. Thus, pushing the IR cut-off to smaller scales alleviates the backreaction constraint.

Dividing  eq.~\eqref{PBnow1} by the above expression, we rewrite the magnetic power spectrum today as,
\begin{eqnarray}
\mcP_B^{\rm now} &\approx& 3\times 10^{-26}\G^2\, \Omega_\em^{\max} (\alpha \xi^2+\beta\xi+\gamma)
\left(\frac{k_i}{1\Mpc^{-1}}\right)^{2}
\left(\frac{T_r}{1\GeV}\right)^{-2}
\left(\frac{g_\ast}{10^2}\right)^{-\frac{2}{3}} \nonumber \\
&& \times
\begin{cases}(a_c/a_r)^2 & (a_c<a_s) \\
a_c a_s/a_r^2 & (a_s<a_c) \,.\\ \end{cases} 
\label{PBnow2}
\end{eqnarray}
Before obtaining the final result for $\mcP_B^{\rm now}$ inserting concrete values of $\xi$ and the conductivity factor, we discuss the third consistency condition from curvature perturbations. 

In the flat slicing, the curvature perturbation $\zeta$ is proportional to the energy density perturbation $\delta\rho$. The electromagnetic fields in our model contributes to   its three  components, $\delta\rho_{\rm em}$, $\delta\rho_\chi$  and $\delta\rho_\phi$ in different ways. 
First, $\delta\rho_{\rm em}$ is nothing but the energy density of the electromagnetic fields themselves. Second, since the spectator scalar field $\chi$ is directly coupled to the electromagnetic fields, its perturbation $\delta\chi$ and hence its density perturbation $\delta\rho_\chi\propto \delta\chi$ are induced. Third, although the inflaton $\phi$ has no direct coupling to $\chi$ or $A_\mu$, they are always coupled gravitationally and thus $\delta\phi$ and $\delta\rho_\phi\propto \delta\phi$ are affected as well.  
In Ref.~\cite{Fujita:2016qab}, all these contributions were computed in a setup similar  to ours which has the same action eq.~\eqref{model lagrangian} except for the last term. There it has been shown that the constraint from the curvature power spectrum is not stronger than the backreaction condition $\Omega_{\rm em }<10^{-1}$, if the IR cut-off is on sufficiently small scale, $k_i\gtrsim 1\Mpc^{-1}$~\cite{Fujita:2016qab}.
Since the dominant contribution to the curvature perturbation on the CMB scale from the electromagnetic fields is generated on  super-horizon scales at which the $\phi F\tilde{F}$ term does not play an important role, we can apply the same argument as Ref.~\cite{Fujita:2016qab} to our case.
Therefore, we focus on the backreaction condition henceforth.

\subsection{The present magnetic field strength}

$\mcP_B^{\rm now}$ in eq.~\eqref{PBnow2} still depends on $a_s/a_r$, $a_c/a_r$ and $\xi$. We discuss the dependence of the scale factors based on the results of Sec.~\ref{When the conductivity stops magnetogenesis} in which $\sigma_c\propto a^{-3/8}$ was studied.

Considering the strong dependence of $\Omega_\em^{\max}$ on $\rho_{\inf}$,
namely $\Omega_\em^{\max}\propto \rho_{\inf}^3$, one cannot strongly increase $\rho_\inf^{1/4}$ from $10^5\GeV$ in order to satisfy the backreaction consistency condition. Since the background energy density $\rho$ decreases as $a^{-3}$ during the inflaton oscillation phase, the scale factor increases during this period by a factor of $a_r/a_e \simeq 10^7\,T_{1\GeV}^{-4/3}\,(\rho_{\inf}^{1/4}/10^5\GeV)^{4/3}$. This factor is much smaller than the value given in eq.~\eqref{aras relation} needed for conductivity not to becomes significant before the end of magnetogenesis, $\eta_s$.
From eq.~\eqref{ac result}, we see that conductivity becomes significant before the stop of the kinetic function $I$ for the present parameter choice and we always have the following hierarchy of the scale factors, $a_c<a_s<a_r$.
Then it is optimal for magnetogenesis to assume that $I$ terminates at the almost same time as reheating,
\begin{equation}
\frac{a_s}{a_r}\simeq 1
\quad\Longrightarrow\quad
\frac{a_c}{a_s}=2\times 10^{-3}T_{\rm 1GeV}^{\frac{8}{57}},
\end{equation}
where we used the first line of eq.~\eqref{ac result}.
Inserting the above scale factors in eq.~\eqref{PBnow2}, we obtain
\begin{align}
\mcP_B^{\rm now}&\approx 10^{-31}\G^2\, \Omega_\em^{\max} (\alpha \xi^2+\beta\xi+\gamma)
\left(\frac{k_i}{1\Mpc^{-1}}\right)^{2}
\left(\frac{T_r}{1\GeV}\right)^{-\frac{98}{57}},
\label{PB last}
\\
\Omega_\em^{\max} &\approx 3\times 10^{-16}\frac{\mathcal{I(\xi)} \sinh(\pi\xi)}{4\xi+5\xi^3+\xi^5}
\left(\frac{\rho_\inf^{1/4}}{10^{5}\GeV}\right)^{12}\left(\frac{k_i}{1\Mpc^{-1}}\right)^{-2}\left(\frac{T_r}{1\GeV}\right)^{-6}.
\label{Omega last}
\end{align}
where we suppress the dependences on $c_\sigma$ and $g_*$.

Eq.~\eqref{PB last} has a positive power-law factor of $\xi$, namely $\alpha \xi^2+\beta\xi+\gamma$, because $\Omega_\em$ has a slightly weaker dependence on $\xi$ than $\mcP_B^{\rm now}$ (see the discussion in Appendix~\ref{Numerical Calculation of} for its reason). 
Consequently, a larger $\xi$ can boost the magnetic field strength $\mcP_B^{\rm now}$ for a given $\Omega_\em^{\max}$.
However, one cannot take arbitrarily large $\xi$ due to the backreaction constraint $\Omega_\em^{\max}\ll 1$.
With the fiducial parameters in eq.~\eqref{Omega last}, 
$\Omega_\em^{\max}=10^{-2}$ corresponds to $\xi=7.6$ which gives 
$\alpha\xi^2+\beta\xi+\gamma\approx 3\times10^2$. 
In that case, we obtain the present strength of the scale-invariant magnetic fields as
\begin{equation}
B_{\rm now}\equiv \sqrt{\mcP_B}\approx 6\times 10^{-16}\G, 
\label{final B}
\end{equation}
which satisfies the lower bound from blazar observations~\cite{Taylor:2011bn}.
The parameters that we adopted for the result~\eqref{final B} are
\begin{equation}
n=3,\ \ m=-\frac{3}{8},\ \ \xi=7.6,\ \ \rho_{\inf}^{1/4}=10^5\GeV,\ \ T_r=1\GeV,
\ \ k_i=1\Mpc^{-1},
\end{equation}
as well as $c_\sigma=g_*=100.$
Note that $\mcP_B^{\rm now}(k)$ is scale invariant only for $k_i<k<k_T$. On larger scale $\mcP_B(k\ll k_i)\propto k^6$ and it decays due to  turbulent damping for $k\ge k_T$.
The damping scale $\lambda_T\equiv 2\pi/k_T$ at present can be roughly estimated as~\cite{Durrer:2013pga,Banerjee:2004df,Caprini:2009pr}
\begin{equation}
\lambda_T\sim \frac{t_{\rm rec}B_{\rm now}}{a_{\rm rec}^3\sqrt{\rho_{\rm rec}}}\sim10^{-1}{\rm pc}\left(\frac{B_{\rm now}}{10^{-15}\G}\right),
\end{equation}
where the subscript ``rec'' denotes recombination.
Therefore, with the above fiducial parameters, a scale-invariant
 magnetic field is realized on scales from $10^{-7}\Mpc$ to $1\Mpc$.

\section{Summary and Discussion}
\label{Summary and Discussion}
In this paper we have shown that by coupling of the electromagnetic field to a spectator field during inflation, we can generate a scale invariant helical magnetic field which is sufficient not only to seed galactic and cluster fields but also satisfy the lower limit on magnetic fields in voids derived from  observations of blazars~\cite{Taylor:2011bn}. To achieve this it is important that the spectator fields keeps rolling during reheating, nearly until the end. Furthermore, the production of charged particles  during reheating leads to the generation of conductivity, which is soon high enough to terminate further magnetogenesis. As long as $I(\chi)$ is large and we normalize it to $1$ at the end of magnetogenesis, the coupling of charged particles is very weak and so are the effects of conductivity. Nevertheless, for the parameter values studied here, it always is the relevant phenomenon to terminate magnetogenesis before $\eta_s$.

In this paper, we did not investigate several other phenomena related to primordial magnetogenesis which might potentially be important. Here we briefly discuss the Schwinger effect, chiral anomaly, and baryogenesis. 

The Schwinger effect is the fact that a electric field which is stronger than a certain critical value (e.g. $E_c=m_e^2/e$ in Minkowski spacetime) decays into pairs of charged particle and anti-particle due to  non-perturbative effects of QED~\cite{Schwinger:1951nm}.
Recently, the Schwinger effect in the expanding universe and its influence on magnetogenesis have been intensively studied~\cite{Frob:2014zka,Kobayashi:2014zza,Hayashinaka:2016qqn,Kitamoto:2018htg,Sobol:2018djj,Banyeres:2018aax}.
Since  strong electric fields are produced during the inflaton oscillation era in our scenario, the Schwinger effect might be relevant. On the other hand, as discussed in Sec.~\ref{Electric Conductivity}, the electric fields are damped efficiently by the high electric conductivity when the kinetic coupling $I$ is still much larger than unity, $I(\eta_c)\simeq 10^8$, in the case of our fiducial parameters. The large value of $I$ effectively suppresses the electromagnetic interaction and hence the critical value of the electric field becomes much higher. Indeed, it was shown in Ref.~\cite{Sobol:2018djj} that the Schwinger effect is negligible until $I$ is close to unity, however, the authors studied the case where magnetogenesis was driven by the inflaton.
In principle, the electric conductivity can be increased by the charged particle production by the Schwinger effect and it would be interesting to explore this possibilities based on our present work.

Recently, it was pointed out that the chiral anomaly, $\partial_\mu J_5^\mu+(e/4\pi)^2 F_{\mu\nu}\tilde{F}^{\mu\nu}=0$ with $J_5^\mu=\bar{\psi}\gamma^\mu\gamma^5\psi$
being the axial current of a massless charged fermion $\psi$, plays an important role in inflationary models with the $U(1)$ Chern-Simons term $\varphi F_{\mu\nu}\tilde{F}^{\mu\nu}$~\cite{Domcke:2018eki}.
The authors of Ref.~\cite{Domcke:2018eki} have explicitly shown that when helical electromagnetic fields are generated by the Chern-Simons term during inflation, the chiral asymmetry of charged fermions should also be produced such that the above chiral anomaly equation holds.
The produced chiral fermions could have a significant impact on the subsequent time evolution of the magnetic fields through the chiral magnetic effect~\cite{Fukushima:2008xe,Boyarsky:2011uy,Akamatsu:2013pjd,Schober:2017cdw,Schober:2018ojn}.
Nevertheless, if the fermions have non-negligible mass, the chiral anomaly equation is modified and the production of the chiral asymmetry can be suppressed.
In particular, for a fermion mass much larger than the Hubble scale during inflation $H_{\inf}\simeq 2{\rm eV}\, (\rho_{\inf}^{1/4}/10^{5}\GeV)^2$, the fermion production is highly suppressed~\cite{Adshead:2018oaa}.
In addition, since the large $I$ effectively reduces the electromagnetic coupling $e$ into $e/I$, the fermion production may be further suppressed.
Therefore, we expect that the production of the chiral asymmetry can be ignored in the parameter region relevant for our scenario.

Even if the chiral asymmetry is not generated during inflation,
helical magnetic fields can cause another fascinating phenomenon around the electroweak phase transition through the chiral anomaly:
The observed baryon asymmetry of our universe can be generated by (hyper-)magnetic fields, if the helicity is negative and the strength is appropriate~\cite{Giovannini:1997eg,Fujita:2016igl,Kamada:2016eeb,Kamada:2016cnb,Jimenez:2017cdr,Kamada:2018tcs}.
Unfortunately, however, we cannot easily apply the results of  previous works to our model, mainly because the kinetic coupling $I$ has not yet become unity at the electroweak phase transition $T\simeq 10^2\GeV$ in our case. Another difference to existing works is that the helical magnetic field does not undergo the inverse cascade process due to its large correlation length. Thus a dedicated investigation is needed to determine the baryon asymmetry produced in our case. We leave it for a future project.

\acknowledgments

We would like to thank Valerie Domcke, Kyohei Mukaida, Ryo Namba and Jennifer Schober for useful comments.
TF is in part supported by the Grant-in-Aid for JSPS Research Fellow No.~17J09103
and the young researchers’ exchange programme between Japan and Switzerland 2018. RD is supported by the Swiss National Science Foundation.

\appendix

\section{Derivation of the Electric Conductivity}
\label{Derivation of the conductivity}

Here, we derive the electric conductivity $\sigma_{c}$  by extending the argument in Ref.~\cite{Caprini:2009yp}.
The EoM for a charged particle with Lorentz force  is
\begin{equation}
m_e\frac{\dd u^\mu}{\dd \tau}=eF^{\mu\nu}u_\nu,
\end{equation}
where $m_e, e$ and $u^\mu=\gamma(1, \bm{v})$ are mass, 4-velocity and charge of the particle, and  $\gamma\equiv 1/\sqrt{1-v^2}$ is the Lorentz factor.
Averaging this equation over a fluid element which contains many charged particles with random velocity directions, one finds that the terms proportional to the velocity vector becomes negligible. However, since the acceleration due to the homogeneous electric field remains, we obtain%
\footnote{Interestingly, the kinetic function $I$ does not explicitly appear in this equation, although $I$ can be interpreted as the modification of the effective charge $e/I$ for the canonical field $IA_\mu$. This is because the effective electric field is $I \bm{E}$ in our notation (e.g. see eq.~\eqref{P of EB}) and hence the combination $e\bm{E}$ is invariant. }
\begin{equation}
\frac{\dd \bar{\bm v}}{\dd t}=\frac{e}{p} \bm{E},
\end{equation}
where $p =\gamma m_e$ is the typical momentum of the particles and $\bar{\bm v}$ denotes the averaged velocity.
Provided that the particles are monotonically accelerated by the electric field for a time $t_a$, the averaged velocity is the order of $\bar{\bm{v}}\simeq t_a e \bm{E}/p$.
Therefore, the averaged current density $\bm{J}$ is expressed in terms of the electric field as
\begin{equation}
\bm{J}\simeq e n \bar{\bm{v}}\simeq t_a\frac{e^2 n}{p}\bm{E}
\quad\Longrightarrow\quad
\sigma_c=t_a\frac{e^2 n}{p},
\label{current eq2}
\end{equation}
where $n$ is the number density of the charged particles.

The acceleration time $t_a$ is often evaluated as the collision time (or the mean free path) which is the inverse of the interaction rate of the charged particles $\Gamma_c$.
However, if the interaction of the charged particles is not significant within the Hubble time, it is more reasonable to estimate $t_a$ as an order of the Hubble time, because the acceleration time cannot be longer than the lifetime of the universe.%
\footnote{The contribution to $\Gamma_c$ from the electromagnetic force is suppressed by $I^4$ in our model. Although the other forces may cause $\Gamma_c>H$ for some particles,
the right-handed leptons which have only the electromagnetic interaction in the standard model might have much smaller $\Gamma_c$ and dominantly contribute to the electric conductivity. We leave the consequence of the different interaction rate between particle species for future work.}
Thus we have
\begin{equation}
t_a\simeq (\max[\Gamma_c, H])^{-1}.
\end{equation}
It should be noted that 
if the interaction rate $\Gamma_c$ is larger than $H$, the particles are expected to be in thermal equilibrium.
In this case $\Gamma_c>H$, using $p\simeq T$ and $t_a\simeq T^2/( e^4 n)$, one finds~\cite{Caprini:2009yp}
\begin{equation}
\sigma_{\rm thermal}\simeq \frac{T}{e^2},
\qquad ({\rm thermal\  equilibrium})
\end{equation}
where $T$ is the temperature of the thermal bath. This expression agrees with eq.~\eqref{sigma100T} aside from the constant numerical factor.
In the other case $H>\Gamma_c$,
the conductivity is given by
\begin{equation}
\sigma_{\rm non-thermal}\simeq \frac{e^2 n}{H p}.
\qquad ({\rm out\ of\  thermal\  equilibrium})
\label{nthermal sigma}
\end{equation}

Since we have already discussed the thermalized case in Sec.~\ref{When the conductivity stops magnetogenesis}, here we consider the case where the electric conductivity is induced by charged particles out of equilibrium.
Again, to compute a concrete time dependence of $\sigma_c$ from eq.~\eqref{nthermal sigma}, a model of the particle production is necessary.
For simplicity, we make the same assumption as eq.~\eqref{rhorad for const Gphi} that inflaton decays into the charged particles
with a constant decay rate $\Gamma_\phi$. 
Note that the difference from the previous case is that the interaction between the decay products are not strong enough to reach thermal equilibrium.
If the inflaton mass is much larger than the charged particle mass, $m_\phi\gg m_e$, the momentum of the produced particle is the order of the inflaton mass $p\simeq m_\phi$. Then, the particle number density is $n\simeq \rho_{\rm rad}/p\simeq \Mpl^2 \Gamma_\phi H/m_\phi$.
Finally using $m_\phi\simeq H_{\inf}$ and $\Gamma_\phi\simeq H_r$, we obtain
\begin{equation}
\sigma_c\simeq \left(\frac{e\Mpl}{H_{\inf}}\right)^2H_r\,.
\end{equation}
Interestingly, the conductivity is constant in this case.
It should be noted that if one considers these charged particles eventually go into thermal equilibrium before the completion of reheating,  
one should shift into eq.~\eqref{sigma100T} at certain time before $\eta_r$, and then the simple ansatz $\sigma_c\propto a^m$ in eq.~\eqref{m introduced}
is not valid.

\section{Numerical Calculation of $\Omega_{\em}$}
\label{Numerical Calculation of}

%
\begin{figure}[tbp]
  \hspace{-2mm}
  \includegraphics[width=70mm]{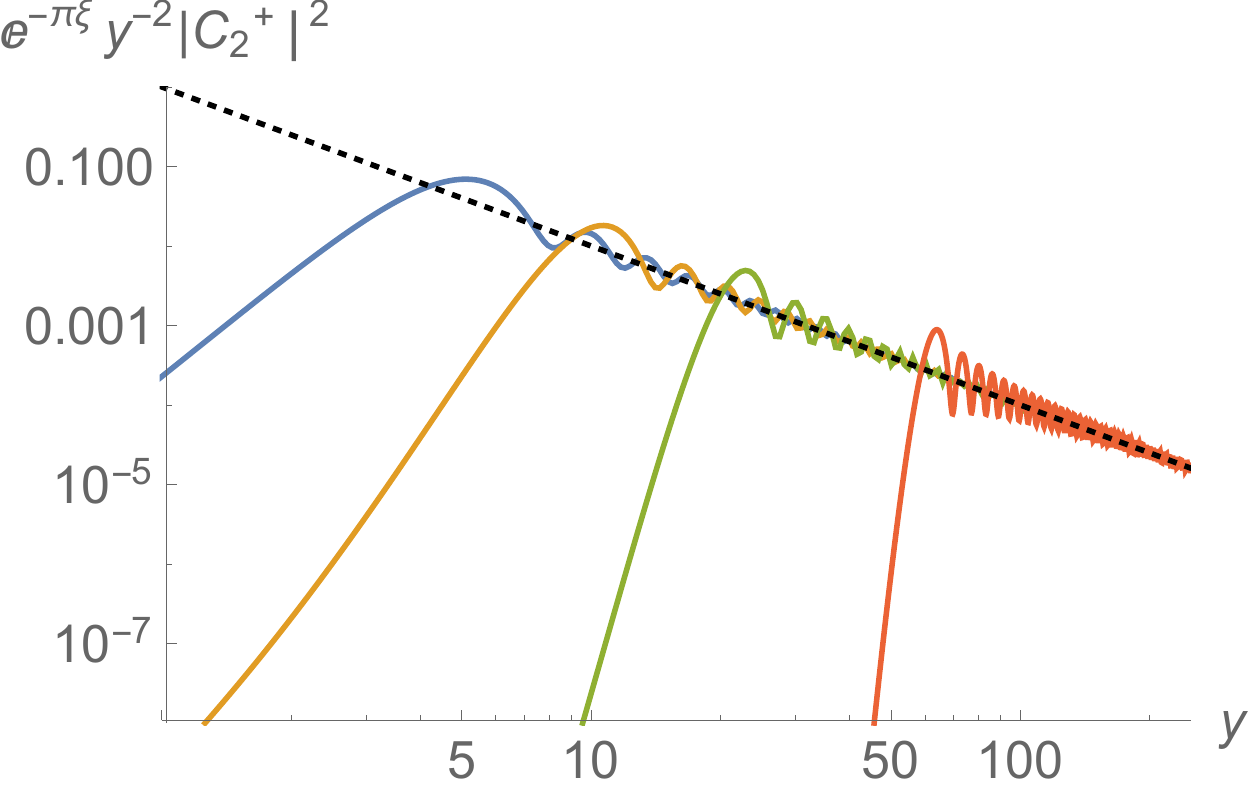}
  \hspace{5mm}
  \includegraphics[width=70mm]{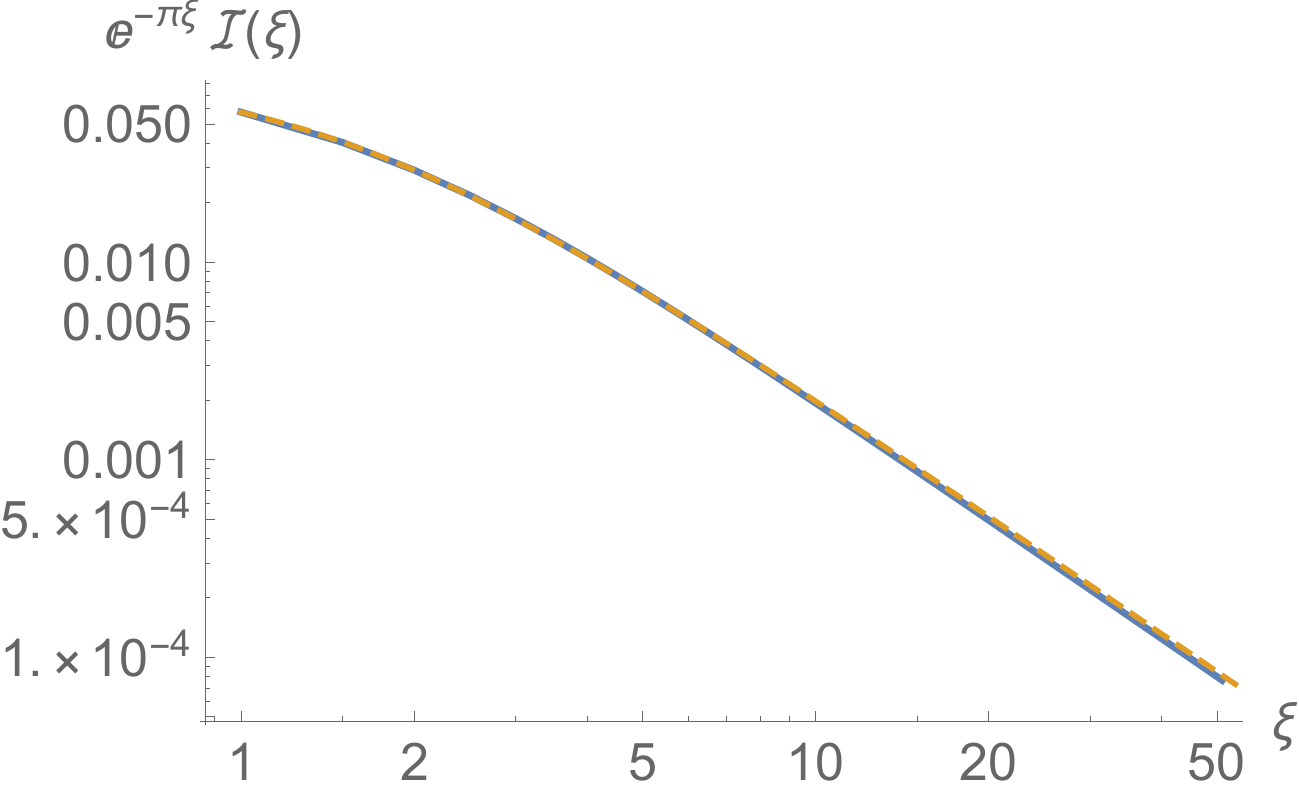}
  \caption
 {{\bf (Left panel)} The integrand of $\mathcal{I}(\xi)$ defined in eq.~\eqref{I integrand} normalized by $e^{-\pi\xi}$ is plotted for $\xi=1$ (blue), $4$ (orange), 10 (green) and 30 (red). The black dashed line denotes $y^{-2}$.
We fix $n=3$. {\bf (Right panel)} The comparison between the numerical integration of $e^{-\pi\xi} \mathcal{I}(\xi)$ (blue thick line) and our fit in eq.~\eqref{I integrand} (orange dashed line). The relative error is within $3\%$ up to $\xi =12$.}
\label{Omega em plot}
\end{figure}
%
To calculate the energy fraction of the electromagnetic fields $\Omega_\em$,
we need to numerically evaluate $\mathcal{I}(\xi)$ defined in eq.~\eqref{I integrand}.
Since we know the asymptotic behavior $|C_2^+(y)|^2= e^{\pi\xi}$ in the limit $y\to \infty$, we are interested in the behavior of $\mathcal{I}(\xi)/e^{\pi\xi}$.
In the left panel of fig.~\ref{Omega em plot}, the integrand of $\mathcal{I}(\xi)$ normalized by $e^{-\pi\xi}$ is shown.
One can see that $e^{-\pi\xi}|C_2^+|^2$ reaches $y^{-2}$ around $y \simeq 2\xi+3$ in the case of $n=3$. Approximating $|C_2^+(y)|^2$ by $e^{\pi\xi}\theta(y-2\xi-3)$
with $\theta(y)$ being the Heaviside function, we obtain
\begin{equation}
\int_0^\infty \frac{\dd y}{y^3}\, \left|C^+_2(y)\right|^2
\sim
e^{\pi\xi}\int_{2\xi+3}^\infty \frac{\dd y}{y^3} = \frac{e^{\pi \xi}}{2(2\xi+3)^2}.
\end{equation}
This expression incentivizes us to fit our numerical result by 
$e^{\pi\xi}/(\alpha\xi^2+\beta\xi+\gamma)$ in eq.~\eqref{I integrand}.
We find $\alpha\approx 4.7,\ \beta\approx2.7,\ \gamma\approx10$ and 
this fit is compared with the numerical integration in the right panel of fig.~\ref{Omega em plot}.

The reason why $y\equiv -k\eta_i\simeq2\xi$ is a threshold
can be seen in eq.~\eqref{inf EoM}. At $-k\eta_i=2\xi$, the sign of $k^2+2\xi k/\eta_i$ flips, and hence the $(+)$-helicity mode gets destabilized and amplified.
As $\xi$ is larger, the lowest $k$-mode which becomes unstable at $\eta=\eta_i$
becomes higher. It pushes the IR cutoff to a higher $k$-mode (i.e. smaller scale).
Therefore $\Omega_\em$ mainly contributed by the IR cutoff mode
relatively decreases as $\xi$ increases, while the amplitude of the scale-invariant part of $\mcP_B^{\rm now}$ does not change.


\end{document}